\documentclass[aps,prd,floatfix,nofootinbib,superscriptaddress,twocolumn,showkeys,10pt]{revtex4-2}
\usepackage{placeins}

\usepackage{amsmath}%,amssymb}

\usepackage{amssymb}
\allowdisplaybreaks

\usepackage{graphicx}
\usepackage{tikzit}
\tikzstyle{new style 0}=[fill=black, draw=none, shape=circle]

\tikzstyle{dashed edge 1}=[-, dashed]
\tikzstyle{dashed edge 2}=[->, dashed]

\usepackage{float}

\usepackage{makeidx}
\usepackage{color}
\usepackage{amsfonts}
\usepackage[usenames,dvipsnames]{pstricks}

\usepackage{epsfig}
\usepackage{pst-grad} % For gradients
\usepackage{mathtools}
\usepackage{mathrsfs}
\usepackage[hyperfootnotes=false]{hyperref}
\hypersetup{colorlinks=true, linkcolor=magenta, filecolor=cyan, urlcolor=blue}
\usepackage{array}
\hypersetup
{	colorlinks,%
	citecolor=black,%
	linkcolor=black,%
	urlcolor=black,%
}

\usepackage{allpurpose}

\usepackage{flushend}% Align the end of the last page.

\def\slashi#1{\rlap{\sl/}#1}

\usepackage{ulem}
\usepackage{color}

\usepackage{calligra}
\usepackage{amsmath}

\begin{document}

\title{Capture Driven Evolution of Asymmetric Dark Matter in Non-rotating Neutron Stars}

\author{Peiran Liu}
\author{Wenrong Sun}
%\email{peiran\_liu@brown.edu}

\affiliation{Department of Physics, Brown University, Providence, RI 02912, USA}

\date{\today}

\begin{abstract}
We develop a self-consistent framework that connects dark matter capture to the long term accumulation and structural evolution of neutron stars, allowing us to quantify the impact of capture driven asymmetric dark matter on gravitational wave observables over astrophysical timescales. We model cold, non-rotating neutron stars using a three layer polytropic EoS coupled to a dark matter component through the two fluid Tolman Oppenheimer Volkoff equations, and reconstruct the long term evolution using a time dependent dark matter capture formalism. By exploring a wide range of astrophysical conditions, including extreme environments designed to maximize the capture efficiency, we derive conservative upper bounds on the effects of dark matter accumulation over a Hubble time. Even under these deliberately optimistic assumptions, the accumulated dark matter component remains subdominant and induces limited structural modifications. In particular, dark matter accumulation increases the stellar compactness while suppressing the Love number and tidal deformability. Significant evolution occurs only in extreme Galactic center like environments over Hubble time scales, whereas for realistic Galactic or cluster dark matter densities the corresponding deviations remain negligible. We therefore conclude that capture driven dark matter accumulation is unlikely to produce detectable signatures in current or near future gravitational wave observations of neutron stars.
\end{abstract}


\maketitle

\section{Introduction}

Dark matter (DM) accumulation in neutron stars (NSs) has been widely studied as a probe of both DM properties and NS microphysics~\cite{Kouvaris:2007ay,bertone2008compact,kouvaris2010can,Leung:2022wcf}. The presence of a dark component can modify the stellar structure and tidal properties, including the compactness \(C\), Love number \(k_2\), and tidal deformability \(\Lambda\)~\cite{hinderer2008tidal,Damour:2009vw,Yagi:2015pkc}. Since these quantities affect the gravitational wave signals emitted during compact binary inspirals, as demonstrated by observations from LIGO/Virgo/KAGRA~\cite{abbott2017gw170817,abbott2018gw170817,LIGOScientific:2021qlt,Radice:2020ddv}, DM admixed neutron stars (DANSs) provide a potential observational window into the dark sector. This motivates the development of a self-consistent framework connecting DM capture, long term accumulation, NS evolution, and gravitational wave observables, thereby enabling a quantitative assessment of the observational consequences of capture driven DM accumulation.

Despite the extensive literature on DANSs~\cite{Leung:2011zz,Leung:2022wcf,Fan:2012qy,ellis2018dark}, the long term evolution of such systems—particularly for asymmetric fermionic dark matter (ADM), which does not annihilate—remains poorly understood. Most existing studies treat the dark component as a static equilibrium configuration and do not connect the capture history of DM to the subsequent evolution of NS structure. The coupled effects of continuous capture, thermalization~\cite{kouvaris2010can,bell2020improved,giorgio2024thermalization}, and the resulting gravitational backreaction are therefore typically neglected. Consequently, it remains unclear whether realistic ADM accumulation histories can produce sufficient DM accumulation to generate observable modifications of NS properties over astrophysical timescales.

Since NSs continuously capture DM throughout their lifetime, the accumulated dark component can gradually grow over cosmological timescales and modify NS observables such as the compactness, Love number, tidal deformability, and stellar stability. This is particularly relevant for ADM, whose non-annihilating nature permits sustained accumulation over a Hubble time rather than a capture annihilation equilibrium. Determining how much DM can realistically accumulate inside NSs, and quantifying the largest structural modifications that such accumulation can induce, therefore provides an important consistency test for DM effects in compact objects and their possible gravitational wave signatures.

In this work, we construct a self-consistent framework for the time dependent accumulation of ADM inside NSs. We model a cold, non-rotating NS using a three layer polytropic equation of state (EoS)~\cite{read2009constraints}, coupled to a dark component within a two fluid Tolman Oppenheimer Volkoff (TOV) formalism~\cite{Leung:2011zz,Leung:2022wcf}. Assuming rapid thermalization of the captured DM, the system evolves through a sequence of quasi-static hydrostatic equilibria. This framework connects DM capture, long term accumulation, stellar evolution, and tidal observables, enabling realistic predictions for the cumulative effects of ADM over astrophysical timescales.

\par
This paper is organized as follows. Sec.~\ref{secnsmodel} presents the three layer NS model, the EoS construction, and the two fluid TOV framework. Sec.~\ref{secDMdistri} introduces the DM capture formalism and the quasi-static time evolution framework, and presents numerical results for different compactnesses, DM masses, and background halo conditions. Sec.~\ref{secnsproperty} studies the time evolution of the Love number and tidal deformability, including the dependence on the EoS and the initial compactness. Sec.~\ref{secconclusion} summarizes the main results and discusses future directions.

Throughout this paper, we use geometrized units with \(G=c=1\).

\section{Neutron Star Structure Model}\label{secnsmodel}

To study the long term structural response of NSs to DM accumulation, we consider cold, non-rotating stellar configurations. 

Over the astrophysical timescales considered in this work, NSs have already cooled far below the proto-NS stage and evolved into quasi-static hydrostatic equilibrium. In this regime, thermal corrections to the stellar EoS are subdominant, while colder NSs more efficiently retain accumulated DM.

We therefore model the NS as a static perfect fluid described by a layered polytropic EoS. The layered construction allows different radial regions of the star to possess different effective stiffness, mimicking the distinction between the dense inner core and the outer stellar layers while maintaining numerical simplicity and flexibility.

The energy momentum tensor in each layer takes the form
\begin{align}
    T^{\mu\nu}_i
    =
    (\rho_i+p_i)u^\mu u^\nu
    +
    p_i g^{\mu\nu}.
\end{align}
At the interfaces between adjacent layers, the pressure and metric functions are required to remain continuous in order to preserve hydrostatic equilibrium.

The spacetime is assumed to be static and spherically symmetric (SSS), with metric
\begin{align}
    ds^2 = -A(r)\,dt^2 + B(r)\,dr^2 + r^2 d\Omega^2,
\end{align}
where
\begin{align}
    A(r) = e^{2\Phi(r)}, \qquad
    B(r) = \left(1 - \frac{2m(r)}{r}\right)^{-1}.
\end{align}

Combining the Einstein field equations with energy momentum conservation,
\begin{align}
    \nabla_{\mu}T^{\mu\nu}=0,
\end{align}
yields the TOV equations,
\begin{subequations}
\begin{align}
    \frac{dp}{dr}&=-(\rho+p)\frac{m+4\pi r^3 p}{r(r-2m)},\label{eqtov1}\\
    \frac{dm}{dr}&=4\pi r^2\rho,\label{eqtov2}\\
    \frac{d\Phi}{dr}&=\frac{m+4\pi r^3p}{r(r-2m)}.\label{eqtov3}
\end{align}
\end{subequations}

These equations define the baseline stellar structure used throughout this work. In the multi layer construction, the EoS is specified piecewise, with different polytropic relations applied in different density regimes. The TOV equations are solved using the local pressure and energy density at each radius, while the enclosed mass function $m(r)$ consistently includes contributions from all interior layers.

When additional components are introduced, the source terms entering Eqs.~\eqref{eqtov1} and \eqref{eqtov3} must be generalized to include the total pressure and energy density of all relevant constituents. 

To close the system, an EoS is required. The NS is modeled using a three layer piecewise polytropic EoS, following \cite{read2009constraints},
\begin{align}\label{eqEOSNS}
    p(r)=\begin{cases}
        K_1\rho^{\Gamma_1},~~~\rho\leq\rho_1,\\
        K_2\rho^{\Gamma_2},~~~\rho_1<\rho\leq\rho_2,\\
        K_3\rho^{\Gamma_3},~~~\rho_2<\rho.
    \end{cases}
\end{align}
where $\rho_1=10^{14.7}g/cm^3=1.85\rho_{nuc},~\rho_2=10^{15.0}g/cm^3=3.70\rho_{nuc}$, $\rho_{nuc}=2.7\times10^{14}g/cm^3$ and $1.4<\Gamma_1<5.0,~1.0<\Gamma_2<5.0,~1.0<\Gamma_3<5.0$. \(K_i\) and \(\Gamma_i\) are the polytropic constants and adiabatic indices, respectively. Here \(\rho\) denotes the mass energy density. The density breakpoints \(\rho_1\) and \(\rho_2\) separate different density regimes within the NS. The three segments correspond approximately to the outer region, intermediate region, and dense core, each characterized by a distinct polytropic index \(\Gamma_i\) that captures changes in the effective stiffness of dense matter. The polytropic constants \(K_i\) are fixed by requiring continuity of the pressure across the matching densities,
\begin{align}
    K_{i+1}
    =
    \frac{p(\rho_i)}{\rho_i^{\Gamma_{i+1}}}.
\end{align}
%From dimensional considerations, the pressure and density scale as $[p] = \mathrm{g\,cm^{-1}\,s^{-2}}$ and $[\rho] = \mathrm{g\,cm^{-3}}$, respectively, implying that the polytropic constant has units
%\begin{align}
%    [K_i] = \mathrm{g}^{\,1-\Gamma_i}\,\mathrm{cm}^{\,3\Gamma_i-1}\,\mathrm{s}^{-2}.
%\end{align}

\par
We consider NS configurations with typical masses $M \sim 1.4\,M_{\odot}$ and radii $R \sim 10\!-\!13~\mathrm{km}$. To examine the dependence of our results on the nuclear EoS, we adopt a representative set of widely used EoS models~\cite{read2009constraints}, summarized in Table~\ref{tabeos}.

\begin{table}[htp!]
    \centering
    \begin{tabular}{l|c|c|c|c|c}
    \hline\hline
    EoS & $\Gamma_1$ & $\Gamma_2$ & $\Gamma_3$ & $M_{\rm max}/M_\odot$ & $R_{1.4}/\mathrm{km}$\\
    \hline\hline
    SLy  & 2.851 & 2.988 & 3.005 & 2.049 & 11.736\\
    \hline
    AP4  & 3.348 & 3.445 & 2.830 & 2.213 & 11.428\\
    \hline
    MPA1 & 2.887 & 3.572 & 3.446 & 2.461 & 12.473\\
    \hline
    H4   & 2.144 & 2.246 & 2.909 & 2.032 & 13.774\\
    \hline\hline
    \end{tabular}
    \caption{
    Representative piecewise polytropic NS EoS adopted from Ref.~\cite{read2009constraints}. The table lists the polytropic indices \(\Gamma_i\), together with the corresponding maximum mass \(M_{\rm max}\) and the radius \(R_{1.4}\) of a canonical \(1.4\,M_\odot\) NS.
    }
    \label{tabeos}
\end{table}
These EoS models span a representative range of stiffness, allowing us to test the robustness of our conclusions against variations in the underlying nuclear microphysics.

We fix the normalization of the piecewise polytropic EOS by imposing pressure continuity across the layer interfaces, starting from a low density outer boundary. Specifically, we terminate the stellar model at
\(\rho_{\rm surf}=10^{10}~\mathrm{g\,cm^{-3}}\)~\cite{Beznogov:2016ejn,Beznogov:2021ijc}
and assign the corresponding pressure
\(p_{\rm surf}=2.5\times10^{24}~\mathrm{dyne\,cm^{-2}}\)~\cite{Choudhuri:2001ai,Potekhin:2013qqa}.
This fixes the normalization of the outermost polytropic segment,
\begin{align}
    K_1
    =
    \frac{p_{\rm surf}}{\rho_{\rm surf}^{\Gamma_1}}.
\end{align}
The remaining polytropic constants are determined recursively by imposing continuity of the pressure at the matching densities,
\begin{align}
    K_{i+1}\rho_i^{\Gamma_{i+1}}
    =
    K_i\rho_i^{\Gamma_i},
    \quad
    \Rightarrow
    \quad
    K_{i+1}
    =
    K_i \rho_i^{\Gamma_i-\Gamma_{i+1}}.
\end{align}
Therefore, once the adiabatic indices \(\Gamma_i\) and transition densities \(\rho_i\) are specified, the complete three layer EOS is fully determined.

For the numerical integration, the TOV equations are initialized at a small but finite radius $r_{\min}=10^{-5}~\mathrm{cm}$, rather than at \(r=0\), to avoid the coordinate singularity at the origin. Throughout this work, we adopt a fixed central density
\begin{align}
    \rho_c
    =
    1.05\times10^{15}~\mathrm{g\,cm^{-3}},
\end{align}
which sets the overall mass scale of the NS configurations. The corresponding central boundary condition is approximated by
\begin{align}
    m(r_{\min})
    \simeq
    \frac{4\pi}{3}\rho_c r_{\min}^3.
\end{align}

For the numerical calculations presented here, we adopt the H4 piecewise polytropic parameterization as the reference NS model, together with the surface density cutoff specified above. This procedure yields a NS configuration with
\begin{align}
    R_{\rm ns}=12.09~\mathrm{km},
    \qquad
    M_{\rm ns}=1.447\,M_{\odot},
\end{align}
consistent with typical NS properties~\cite{de2018tidal,lattimer2001neutron}. Because the stellar integration is terminated at the effective surface density
\(\rho_{\rm surf}=10^{10}~\mathrm{g\,cm^{-3}}\), with the low density EoS normalization fixed by the corresponding pressure
\(p_{\rm surf}=2.5\times10^{24}~\mathrm{dyne\,cm^{-2}}\), the resulting radius should not be identified directly with the tabulated \(R_{1.4}\) values in Tab.~\ref{tabeos}. correspond to the full piecewise polytropic EoS fits. Unless otherwise stated, this configuration is adopted as the baseline NS model throughout the following analysis.

\section{Time Evolution of the Dark Matter Component}\label{secDMdistri}

To model the time dependent accumulation of DM inside NSs, we describe the dark component as a degenerate fermion gas. Fermionic DM naturally supports an extended DM core through degeneracy pressure generated by the Pauli exclusion principle. As captured particles accumulate near the stellar core, the resulting degeneracy pressure counteracts gravitational compression, allowing the dark component to be described by an independent EoS within the two fluid framework.

We focus on asymmetric dark matter (ADM) because it naturally allows long term accumulation inside NSs. The suppressed annihilation rate in ADM models permits continuous accumulation over astrophysical timescales and maximizes the total captured DM population. The ADM framework therefore provides a conservative upper limit estimate for the possible structural impact of DM accumulation.

In the ideal degenerate Fermi gas approximation, the DM EoS is described by the standard zero temperature expressions~\cite{panotopoulos2017dark},
\begin{align}
    \rho_{\chi}&=\frac{1}{\pi^2}\int_{0}^{k_{F}}k^2\sqrt{(k\hbar)^2+m_{\chi}^2}\,dk \nonumber\\
    &=\frac{1}{8\hbar^3\pi^2}\left\{k_F\hbar\sqrt{(k_F\hbar)^2+m_{\chi}^2}\left[2(k_F\hbar)^2+m_{\chi}^2\right]\right.\nonumber\\
    &\left.-m_{\chi}^4\mathrm{arcsinh}\left(\frac{k_F\hbar}{m_{\chi}}\right)\right\}, \label{eqdensityDMFer}\\
    p_{\chi}&=\frac{1}{3\pi^2}\int_0^{k_{F}}\frac{k^2(k\hbar)^2}{\sqrt{(k\hbar)^2+m_{\chi}^2}}\,dk \nonumber\\
    &=\frac{1}{24\hbar^3\pi^2}\left\{k_F\hbar\left[2(k_F\hbar)^2-3m_{\chi}^2\right]\sqrt{(k_F\hbar)^2+m_{\chi}^2}\right.\nonumber\\
    &\left.+3m_{\chi}^4\mathrm{arcsinh}\left(\frac{k_F\hbar}{m_{\chi}}\right)\right\}, \label{eqpressureDMFer}
\end{align}
where \(m_\chi\) is the DM particle mass. Throughout this work, we assume a spin degeneracy factor \(g=2\), corresponding to a spin $1/2$ fermionic DM particle. The Fermi momentum is therefore given by
\begin{align}
    k_F = \left(3\pi^2 n_{\chi}\right)^{1/3}.
\end{align}
where \(n_\chi\) is the DM number density. The DM component is therefore treated as a perfect fluid described by this EoS.

\subsection{Formation of the Dark Matter Core}\label{secNSDMcore}

We model the formation of a DM core in the central region of the NS. After capture, DM particles lose kinetic energy through repeated scattering with the stellar medium and gradually thermalize with the surrounding matter. During this early stage, the captured DM forms an approximately thermal distribution characterized by the thermal radius
\begin{align}
    r_{\rm th}
    \sim
    \left(
    \frac{9k_B T_c}
    {4\pi \rho_c m_{\chi}}
    \right)^{1/2},
\end{align}
where \(\rho_c\) and \(T_c\) denote the central density and temperature of the NS, respectively~\cite{bertone2008compact}.

As the captured DM population grows, the central number density and Fermi momentum both increase. Once the degeneracy pressure exceeds the thermal pressure supporting the initial distribution, the Maxwell Boltzmann description breaks down, and the thermal radius no longer provides an adequate characterization of the DM configuration. The onset of the degenerate regime is discussed quantitatively in Subsubsec.~\ref{subsubsecdeg}. The dark component then transitions into a degenerate self-gravitating core described by the fermionic EoS introduced above. The subsequent evolution is therefore modeled within a two fluid TOV framework in which the baryonic and dark components jointly determine the spacetime geometry.

In this regime, the baryonic and DM components are treated as two coupled fluids interacting only through gravity and sharing a common spacetime metric. The hydrostatic equilibrium equations are
\begin{subequations}
\begin{align}
    \frac{dp_{\rm bar}}{dr}
    &=
    -(\rho_{\rm bar}+p_{\rm bar})
    \frac{
    m_{\rm tot}+4\pi r^3 (p_{\rm bar}+p_{\chi})
    }{
    r(r-2m_{\rm tot})
    },
    \label{eqtovdmp1}\\
    \frac{dp_{\chi}}{dr}
    &=
    -(\rho_{\chi}+p_{\chi})
    \frac{
    m_{\rm tot}+4\pi r^3 (p_{\rm bar}+p_{\chi})
    }{
    r(r-2m_{\rm tot})
    },
    \label{eqtovdmp2}\\
    \frac{dm_{\rm tot}}{dr}
    &=
    4\pi r^2(\rho_{\rm bar}+\rho_{\chi}),
    \label{eqtovdmm}\\
    \frac{d\Phi}{dr}
    &=
    \frac{
    m_{\rm tot}+4\pi r^3 (p_{\rm bar}+p_{\chi})
    }{
    r(r-2m_{\rm tot})
    },
    \label{eqtovdmphi}
\end{align}
\end{subequations}
where \(m_{\rm tot}(r)\) denotes the enclosed mass of both components. The DM component is modeled using the degenerate fermion gas EoS given in Eqs.~\eqref{eqdensityDMFer} and \eqref{eqpressureDMFer}.

\subsubsection{Static Dark Matter Configurations}\label{subsubsecQsevoNP}

Because DM capture occurs over astrophysical timescales, whereas the internal dynamical timescale of the NS is extremely short,
\begin{align}
    \tau_{\rm dyn}
    \sim
    1/\sqrt{\rho}
    \sim
    10^{-4}-10^{-3}\,\mathrm{s},
\end{align}
the stellar structure remains close to hydrostatic equilibrium throughout the accumulation process~\cite{shapiro2024black,glendenning2012compact}. This strong separation of timescales justifies the quasi-static approximation. Rather than solving the full time dependent relativistic hydrodynamic system, we construct a sequence of static two fluid TOV solutions describing the gradual accumulation of DM inside the NS.

For each solution, we fix the gravitational mass of the underlying baryonic NS and specify a central DM number density \(n_\chi(0)\). The corresponding baryonic central density is determined numerically to match the target baryonic gravitational mass. The DM pressure and energy density are then obtained from the fermionic EoS given in Eqs.~\eqref{eqdensityDMFer} and \eqref{eqpressureDMFer}, and the coupled two fluid TOV equations are integrated outward from the stellar center. Repeating this procedure for different values of \(n_\chi(0)\) generates a family of NS configurations with increasing DM content.

To isolate the structural effect of the dark component, each solution is compared against the corresponding purely baryonic NS with the same baryonic gravitational mass. This allows us to quantify how accumulated DM modifies global stellar properties such as the stellar radius, compactness, Love number, and tidal deformability.

We solve the coupled two fluid TOV system Eqs.~\eqref{eqtovdmp1}-\eqref{eqtovdmphi} subject to the central boundary conditions
\begin{align}
    m_{\rm tot}(0)=0,
    \qquad
    n_\chi(0)=n_{\chi,c},
    \qquad
    \rho_{\rm bar}(0)=\rho_c.
\end{align}
Equivalently, the solutions may be parametrized by the central Fermi momentum \(k_{F,c}\).

According to the piecewise polytropic EoS in Eq.~\eqref{eqEOSNS}, the baryonic core boundary is defined by
\begin{align}
    \rho_{\rm bar}(R_1)
    =
    \rho_2
    =
    10^{15}~\mathrm{g\,cm^{-3}}.
\end{align}

For the dark component, we define the distribution radius \(R_{\rm DM}\) as the radius at which the DM number density falls below the degeneracy threshold corresponding to the NS core temperature \(T\sim10^7~\mathrm{K}\).
\begin{align}
    n_\chi(R_{\rm DM})
    \simeq
    n_{\rm deg}
    \approx
    9.9\times10^{30}~\mathrm{cm^{-3}},
\end{align}
for the benchmark DM mass \(m_\chi=1~\mathrm{GeV}\). Physically, this defines the radius beyond which the captured DM is no longer degenerate, and the zero temperature Fermi gas EoS adopted in this work ceases to be applicable. We have verified that the resulting stellar properties remain insensitive to moderate variations of this cutoff criterion.

\begin{figure}
    \centering
    \includegraphics[width=0.8\linewidth]{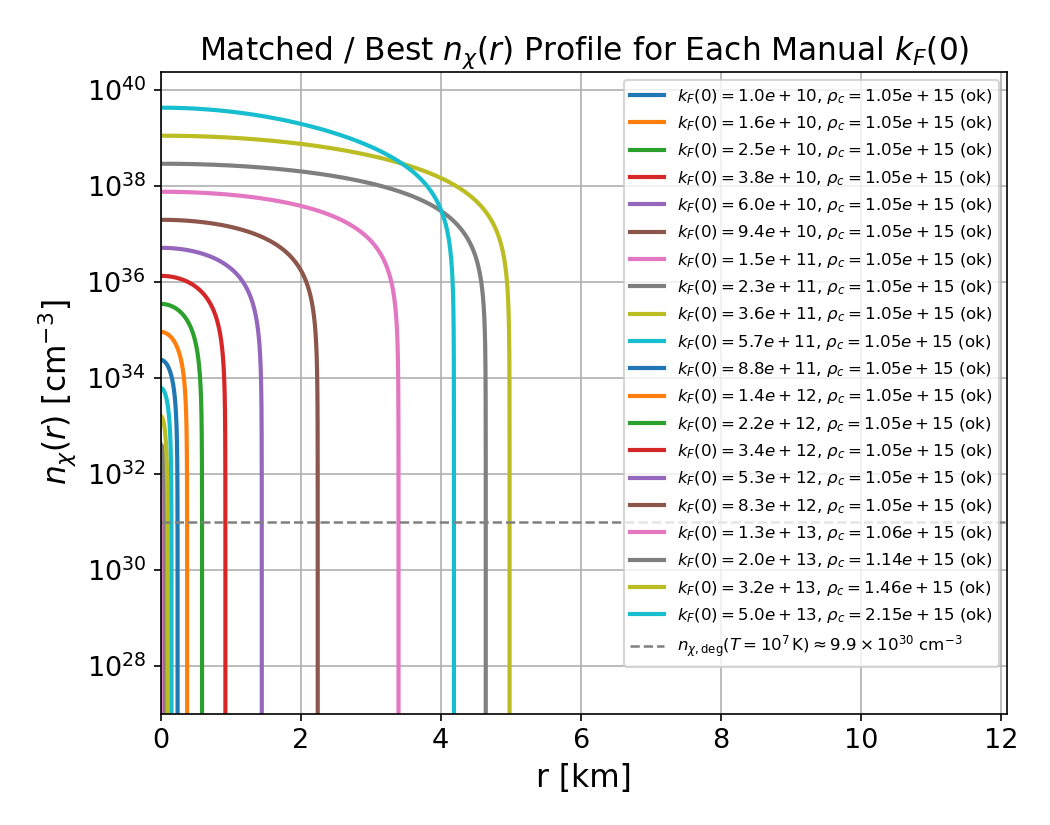}\\
    (a)\\
    \includegraphics[width=0.8\linewidth]{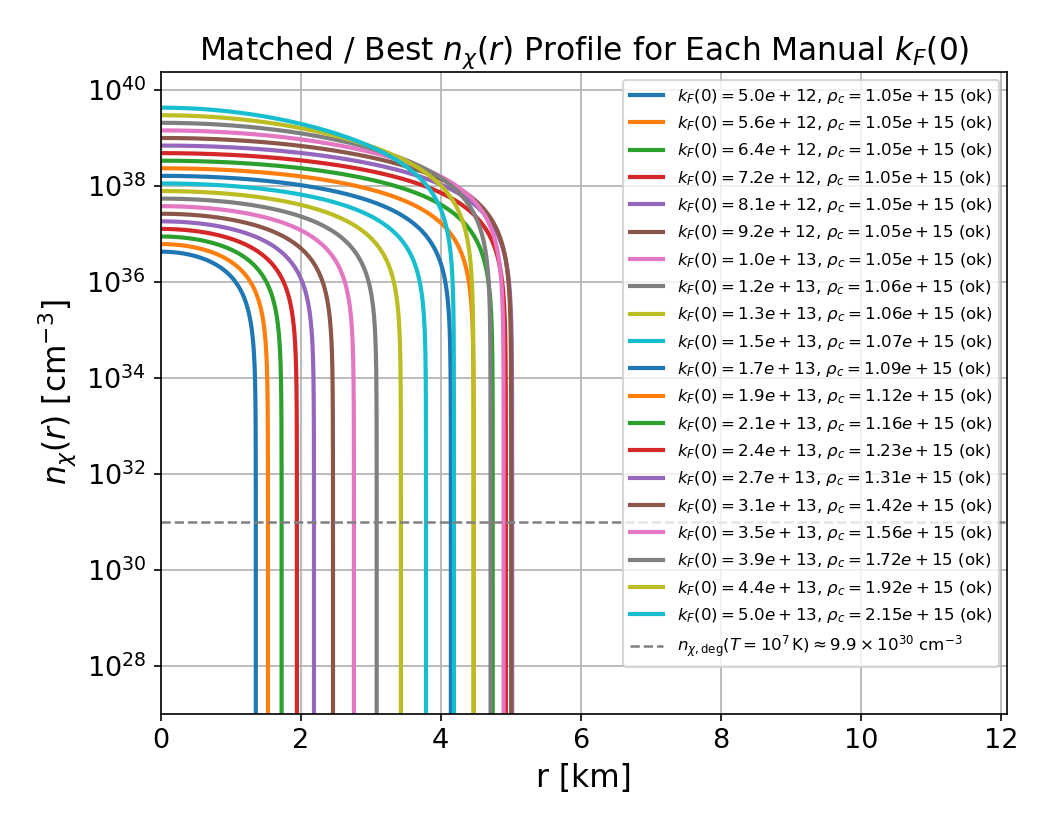}\\
    (b)
    \caption{
Radial DM number density profiles \(n_\chi(r)\) for different central Fermi momenta \(k_{F,c}\), assuming a DM particle mass \(m_\chi=1~\mathrm{GeV}\). (a) shows representative profiles over the full range \(k_{F,c}=10^{10}\!-\!5\times10^{13}\,\mathrm{cm^{-1}}\), (b) focuses on the high-density regime \(k_{F,c}=5\times10^{12}\!-\!5\times10^{13}\,\mathrm{cm^{-1}}\) to illustrate the evolution of the DM profile at large accumulated DM fractions. The horizontal dashed line denotes the degeneracy threshold \(n_{\rm deg}\), whose intersection with each profile defines the characteristic DM distribution radius \(R_{\rm DM}\). As \(k_{F,c}\) increases, both the central DM density and the spatial extent of the degenerate DM core increase.
}
    \label{fignchi}
\end{figure}

From these solutions, we obtain the DM density profiles and the corresponding characteristic radius \(R_{\rm DM}\) as functions of the central Fermi momentum \(k_{F,c}\). This allows us to quantify how the spatial extent of the dark component evolves with increasing DM accumulation. The resulting profiles and the dependence of \(R_{\rm DM}\) on \(k_{F,c}\) are shown in Fig.~\ref{fignchi}.

From these solutions, we obtain the DM density profiles and the corresponding characteristic radius \(R_{\rm DM}\) as functions of the central Fermi momentum \(k_{F,c}\). This allows us to quantify how the spatial extent of the dark component evolves with increasing DM accumulation. The horizontal dashed line in Fig.~\ref{fignchi} marks the degeneracy threshold \(
n_{\rm deg}\simeq9.9\times10^{30}~\mathrm{cm^{-3}},
\) below which the captured DM is no longer described by the zero temperature degenerate Fermi gas EoS adopted in this work. We therefore define the characteristic DM distribution radius \(R_{\rm DM}\) as the radius where the density profile intersects this threshold. The resulting density profiles and the dependence of \(R_{\rm DM}\) on \(k_{F,c}\) are shown in Fig.~\ref{fignchi}.

From these solutions, we obtain the DM density profiles and the corresponding characteristic radius \(R_{\rm DM}\) as functions of the central Fermi momentum \(k_{F,c}\). To highlight the evolution of the DM profile after substantial DM accumulation, we also show representative solutions with relatively large values of \(k_{F,c}\). This allows us to quantify how the spatial extent of the dark component evolves with increasing DM accumulation.

%To reconstruct the time evolution of the dark-matter distribution, we model the capture of dark-matter particles by the neutron star and relate it to the growth of the total accumulated dark-matter number.

\subsection{Dark Matter Capture and Time Evolution}

The long term evolution of the DM population inside the NS is governed by the capture of halo DM particles through scattering with the stellar medium. Since annihilation is negligible in the asymmetric DM scenario considered here, the accumulated particle number evolves according to
\begin{align}
    \frac{dN_\chi}{dt}
    =
    C_{\rm cap},
\end{align}
where \(C_{\rm cap}\) denotes the DM capture rate.

Physically, the capture process occurs when an incoming DM particle loses sufficient kinetic energy through scattering with baryonic matter inside the NS and becomes gravitationally bound to the system. Neglecting Pauli-blocking effects, the capture rate is given by~\cite{bell2020improved,kouvaris2010can,giorgio2024thermalization}
\begin{align}\label{eqcap}
    C_{\rm cap}
    =&
    \frac{4\pi}{v_{\rm ns}}
    \frac{\rho_{\rm halo}}{m_{\chi}}
    \mathrm{Erf}\!\left(
    \sqrt{\frac{3}{2}}
    \frac{v_{\rm ns}}{v_{d}}
    \right)\nonumber\\&\times
    \int_{0}^{R_{\rm ns}}
    r^2 dr \,
    n_b(r)
    \frac{1-A(r)}{A(r)}
    \,
    \langle\sigma(r)\rangle,
\end{align}
with
\begin{align}
    \langle\sigma(r)\rangle
    =
    \frac{a}{16\pi m_{\chi}^2}
    \left[
    \frac{
    4(1-A(r))m_{\chi}
    }{
    A(r)(1+\mu^2)
    }
    \right]^n
    \frac{1}{n+1}.
\end{align}

Here, \(\rho_{\rm halo}\) denotes the ambient DM halo density, \(v_{\rm ns}\) is the NS velocity, \(v_d\) is the DM velocity dispersion, and \(n_b(r)\) is the baryon number density profile inside the NS. The mass ratio is defined by
\begin{align}
    \mu
    =
    \frac{m_{\chi}}{m_n},
\end{align}
where \(m_n\) is the neutron mass. The squared matrix element is parameterized as $|\bar{M}|^2 = a t^n$, where \(t\) is the momentum transfer variable. The cases \(n=0,1,2\) correspond to momentum independent, linearly momentum suppressed, and quadratically momentum suppressed interactions, respectively. In the numerical calculations presented below, we adopt the momentum independent case (\(n=0\)), for which $a=16\pi m_{\chi}^2 \sigma_{\rm ref}$, with a reference cross section $\sigma_{\rm ref}=1.7\times10^{-45}~\mathrm{cm^2}$.

The evaluation of the capture rate requires the baryon number density profile \(n_b(r)\) inside the NS. Assuming zero temperature matter, the baryon number density is obtained from the thermodynamic relation
\begin{align}
    \frac{dn_b}{n_b}
    =
    \frac{d\rho_{\rm bar}}{\rho_{\rm bar}+p_{\rm bar}},
\end{align}
which follows from the first law of thermodynamics for a cold relativistic fluid. Because the NS EoS is defined as a layered polytropic model, this relation is integrated separately within each EoS segment
\begin{align}
    &n_{b,i}(r)
    \nonumber\\&=
    n_{b,i}(R_i)
    \exp\!\left[
    \int_{R_i}^{r}
    \frac{1}{\rho_{\rm bar}(r')+p_{\rm bar}(r')}
    \frac{d\rho_{\rm bar}(r')}{dr'}\,dr'
    \right],
\end{align}
where \(R_i\) denotes the radius corresponding to the transition between adjacent EoS layers.

At the stellar surface, we impose the boundary condition
\begin{align}
    n_b(R_{\rm ns})
    \simeq
    \frac{\rho_{\rm bar}(R_{\rm ns})}{m_n},
\end{align}
where \(m_n\) is the neutron mass.

\begin{figure}[htp!]
    \centering
    \includegraphics[width=0.8\linewidth]{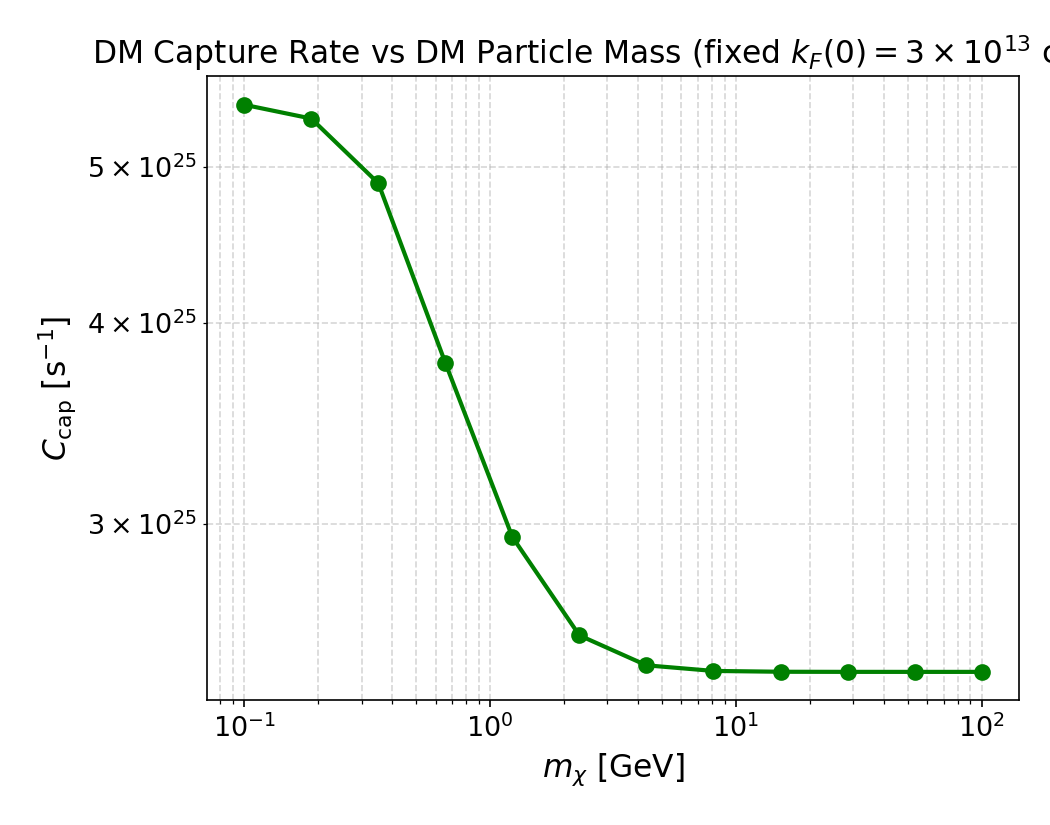}\\
    ~~~~~(a)\\
    \includegraphics[width=0.8\linewidth]{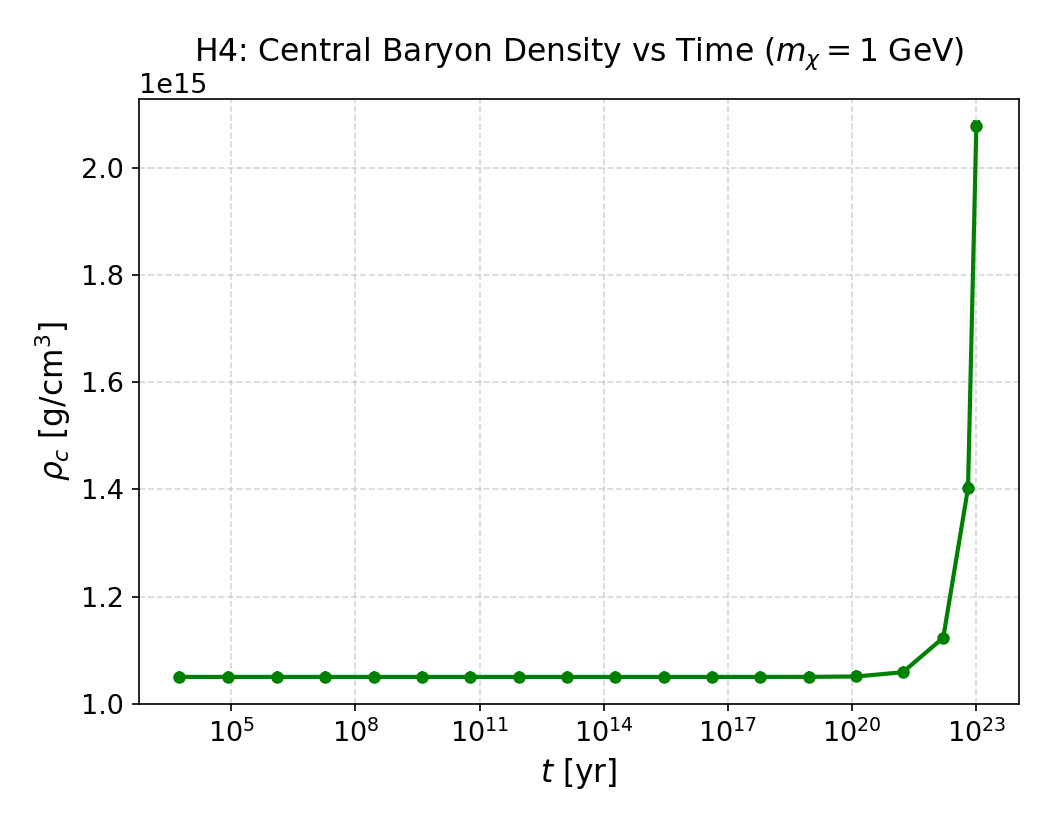}\\
    ~~~~~(b)\\
\caption{
(a) DM capture rate \(C_{\rm cap}\) as a function of the DM particle mass \(m_\chi\). The capture rate increases toward lower DM masses, implying more efficient long term accumulation for lighter DM particles. (b) Reconstructed quasi-static evolution of the central baryon density for the reference H4 NS model during DM accumulation. As the accumulated DM population increases, the additional gravitational contribution of the dark component gradually compresses the baryonic core, leading to a higher central baryon density.
}
    \label{figcaptureratemchi_trhoc}
\end{figure}

To estimate the maximum possible DM accumulation inside the NS, we evaluate the capture rate in Eq.~\eqref{eqcap} for fermionic DM particles in the mass range
\[
0.1~\mathrm{GeV}
\leq
m_\chi
\leq
100~\mathrm{GeV}.
\]
For each choice of \(m_\chi\), the capture integral is computed numerically over the full NS profile from \(r=0\) to \(R_{\rm ns}\)~\cite{zurek2014asymmetric,Kaplan:2009ag,Petraki:2013wwa}. The resulting \(C_{\rm cap}\)--\(m_\chi\) relation is shown in Fig.~\ref{figcaptureratemchi_trhoc}(a), where the capture rate is seen to increase toward lower DM masses. This behavior motivates our focus on light fermionic DM when estimating the maximal possible DM accumulation inside NSs.

As shown in Fig.~\ref{figcaptureratemchi_trhoc}(a), lighter DM particles lead to significantly larger capture rates and therefore more efficient long term accumulation inside the NS.

The resulting quasi-static evolution of the stellar structure is illustrated in Fig.~\ref{figcaptureratemchi_trhoc}(b). Using the capture rate in Eq.~\eqref{eqcap}, we reconstruct the quasi-static evolution by mapping the accumulated DM particle number at each time to the corresponding two fluid TOV equilibrium solution while keeping the baryonic component fixed. As DM accumulates, its additional gravitational contribution gradually compresses the baryonic core, increasing the central baryon density and the stellar compactness. Although the accumulated DM mass remains subdominant compared with the baryonic component, its gravitational backreaction on the NS structure becomes increasingly noticeable at larger accumulated DM fractions.

\subsubsection{Time Evolution}

To reconstruct the time evolution of the DM core, we first determine the accumulated DM particle number from the capture rate through
\begin{align}
    \frac{dN_{\chi}(t)}{dt}
    =
    C_{\rm cap}(t).
\end{align}

As discussed in Subsubsec.~\ref{subsubsecQsevoNP}, we construct a sequence of equilibrium two fluid TOV solutions parameterized by the central Fermi momentum \(k_{F,c}\). For each value of \(k_{F,c}\), the coupled TOV equations determine the corresponding DM number density profile \(n_\chi(r)\). This establishes a mapping between the central Fermi momentum and the total accumulated DM particle number
\begin{align}\label{eqNchi}
    N_{\chi}
    =
    4\pi
    \int_{0}^{R_{\rm DM}}
    r^2 n_{\chi}(r)\,dr.
\end{align}
In practice, the integral is evaluated numerically using trapezoidal summation over the discretized radial profile obtained from the TOV integration.

Because the DM component contributes to the gravitational potential, it modifies the NS metric and therefore alters the capture rate itself. The evolution is thus intrinsically coupled and must be treated self-consistently. For every equilibrium configuration, we therefore compute the corresponding DM distribution, NS metric, total particle number, and capture rate, establishing the mapping
\begin{align}
    k_{F,c}
    \;\longrightarrow\;
    \{
    N_\chi,\;
    C_{\rm cap},\;
    \text{stellar properties}
    \}.
\end{align}

The time evolution is reconstructed by matching the accumulated particle number \(N_\chi(t)\) to the equilibrium sequence. Numerically, we discretize the evolution by comparing neighboring equilibrium configurations and approximate the corresponding time increment as
\begin{align}
    \Delta t
    \simeq
    \frac{\Delta N_{\chi}}{C_{\rm cap}},
\end{align}
where \(C_{\rm cap}\) is evaluated for each configuration. This procedure maps the long term accumulation process onto a sequence of quasi-static equilibrium states describing the coupled evolution of the DM core and the NS.

We present results for two representative DM environments:
(i) a canonical Galactic background density, and
(ii) an extreme high density environment corresponding to dense DM spikes. The results for the canonical Galactic background environment are presented first, followed by the maximized capture scenario in Subsubsec.~\ref{subsubsecmax}.

To illustrate the long term accumulation of DM under typical Galactic conditions, we reconstruct the quasi-static evolution of the DM core using a canonical local DM halo density
\begin{align}
    \rho_{\rm halo}
    =
    0.3\,\mathrm{GeV/cm^3}.
\end{align}

We further adopt NS and DM velocity dispersions
\begin{align}
    v_{\rm ns}
    =
    300~\mathrm{km/s},
    \quad
    v_d
    =
    270~\mathrm{km/s},
\end{align}
corresponding to typical Galactic kinematics for NSs and virialized halo DM.

\begin{figure}[htp!]
    \centering
    \includegraphics[width=0.8\linewidth]{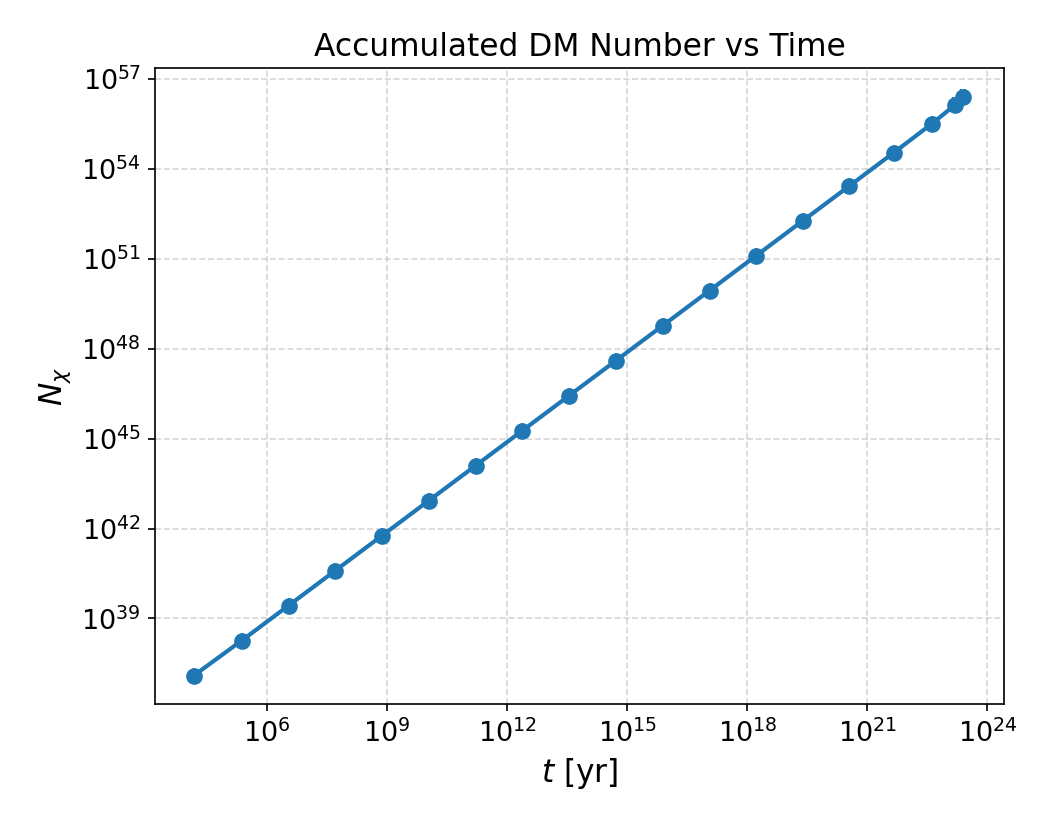}\\
    ~~~~(a)\\
    \includegraphics[width=0.8\linewidth]{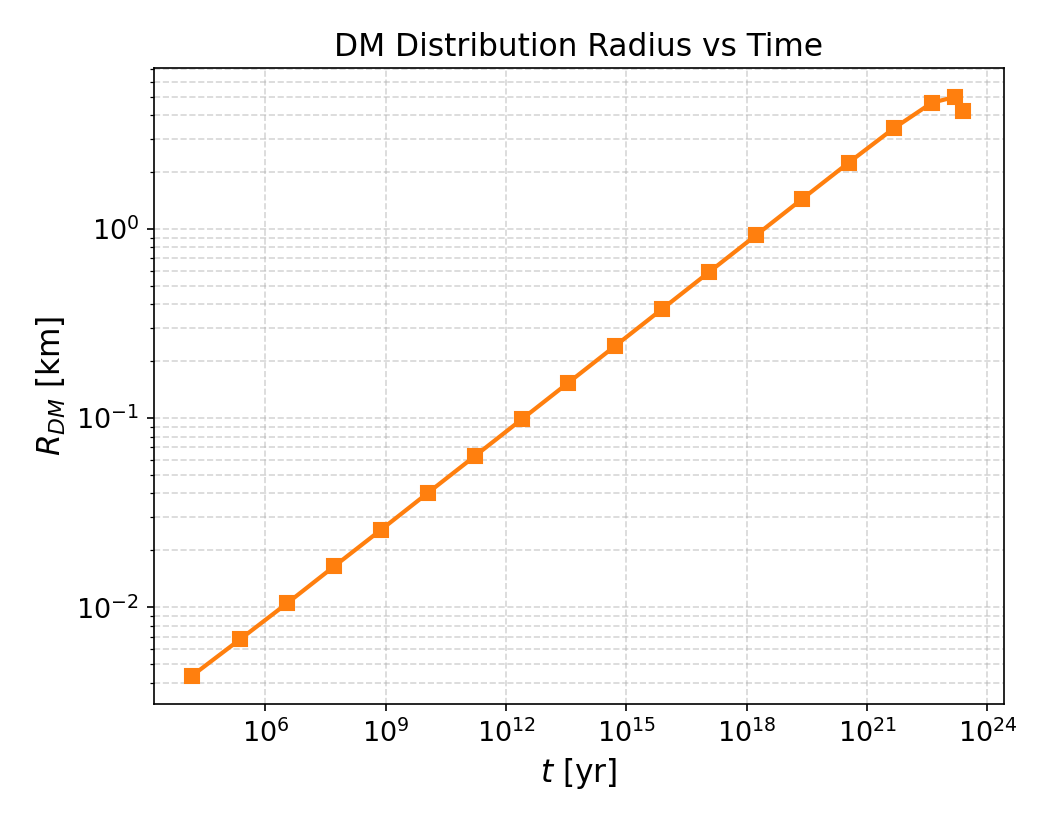}\\
    ~~~~(b)\\
    \includegraphics[width=0.8\linewidth]{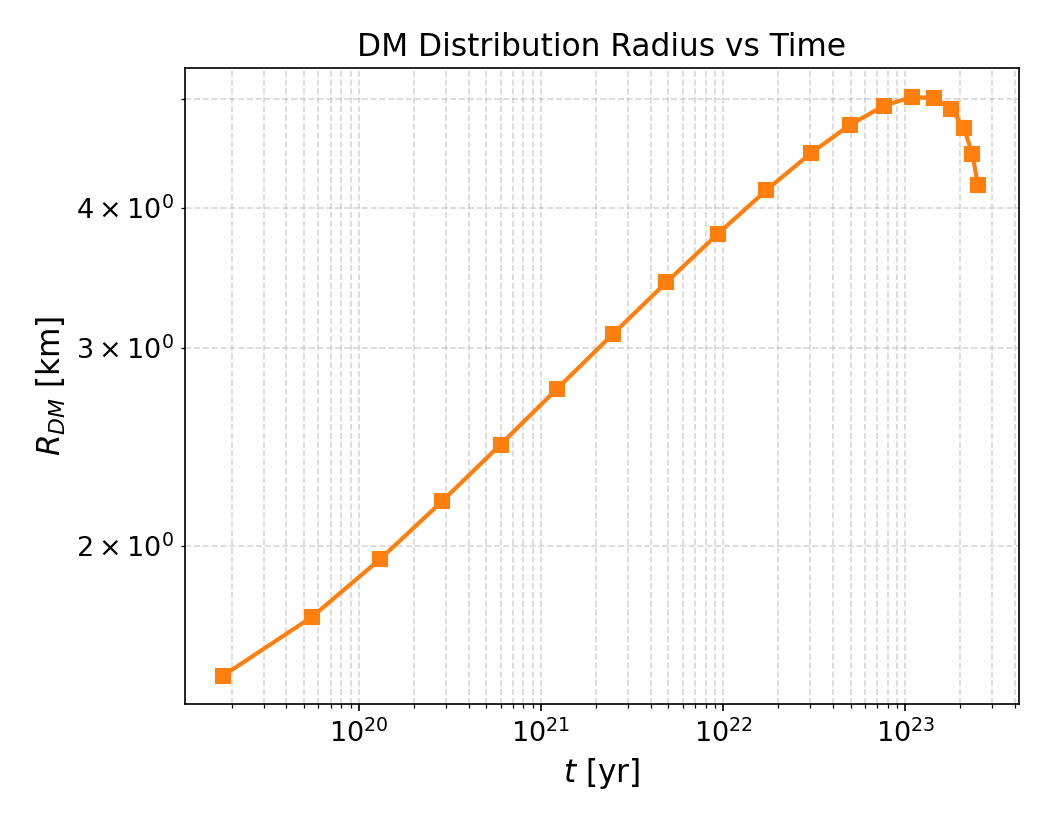}\\
    ~~~~(c)\\
    \caption{
Time evolution of the accumulated DM particle number \(N_\chi\) and the characteristic DM distribution radius \(R_{\rm DM}\) for a canonical Galactic DM halo density \(\rho_{\rm halo}=0.3\,\mathrm{GeV/cm^3}\). (a) Evolution of the accumulated DM particle number. (b) Evolution of the characteristic DM distribution radius. (c) Zoomed in view of (b) over \(10^{19}\!-\!10^{23}\,\mathrm{yr}\), highlighting the late time decrease of \(R_{\rm DM}\).
}
    \label{figtRDMH4}
\end{figure}

Using these parameters, we reconstruct the time evolution of the total accumulated DM particle number \(N_{\chi}\) and the characteristic DM distribution radius \(R_{\rm DM}\), as shown in Fig.~\ref{figtRDMH4}. As expected, \(N_{\chi}\) increases monotonically with time due to continuous capture. The characteristic radius \(R_{\rm DM}\) shown in Fig.~\ref{figtRDMH4}(b), exhibits a non-monotonic behavior that will be discussed below.

An interesting feature of Fig.~\ref{figtRDMH4}(b) is that the characteristic DM radius $R_{\rm DM}$ does not increase monotonically throughout the accumulation history. Instead, after reaching a maximum, it undergoes a slight decrease at very late times. To examine this behavior more clearly, we compute a denser sequence of equilibrium configurations in the high central density regime, as shown in Fig.~\ref{figtRDMH4} (c). The resulting evolution confirms that the late time decrease of $R_{\rm DM}$ is a robust feature of the equilibrium sequence. Physically, this behavior reflects the competition between the continued accumulation of DM and the increasing self-gravity of the accumulated dark component. At early times, the growth of the DM population causes the spatial extent of the degenerate core to expand. At sufficiently large central densities, however, the self-gravity of the dark component becomes increasingly important and gradually compresses the dark core, leading to a slight reduction of $R_{\rm DM}$.

\subsubsection{Maximized Capture Scenario}\label{subsubsecmax}

To estimate the maximum DM accumulation achievable within a Hubble time, we consider an intentionally extreme astrophysical environment designed to maximize the capture rate in Eq.~\eqref{eqcap}. The purpose of this setup is not to model a typical NS environment, but rather to establish a conservative upper bound on the structural modifications that capture driven DM accumulation can produce over a Hubble time.

In particular, we adopt a very large ambient DM density
\begin{align}
    \rho_{\rm halo}
    =
    2.2\times10^{9}\,M_{\odot}/\mathrm{pc}^3,
\end{align}
corresponding to dense DM spikes that may arise in extreme galactic center environments~\cite{shen2024exploring}. We further maximize the velocity dependent factor in the capture rate by taking
\begin{align}
    \mathrm{Erf}\!\left(
    \sqrt{\frac{3}{2}}
    \frac{v_{\rm ns}}{v_d}
    \right)
    \approx
    1,
\end{align}
which corresponds to the limit \(v_{\rm ns}\gg v_d\).

In realistic astrophysical systems, however, regions with large DM densities are typically associated with deep gravitational potentials and therefore larger NS velocities, while slowly moving NSs are more commonly found in lower density environments. This introduces a natural competition between density enhancement and velocity suppression in the capture process.

We assume that the NS remains in such an extreme environment throughout the entire capture history and adopt a representative low NS velocity
\begin{align}
    v_{\rm ns}
    =
    30\,\mathrm{km/s},
\end{align}
consistent with the lower end of observed NS velocity distributions~\cite{hobbs2005statistical,faucher2006birth,Disberg:2025xoh}. Combined with the high density assumption above, this choice specifies the benchmark parameters adopted for the maximized capture scenario.

\begin{figure}[htp!]
    \centering
    \includegraphics[width=0.8\linewidth]{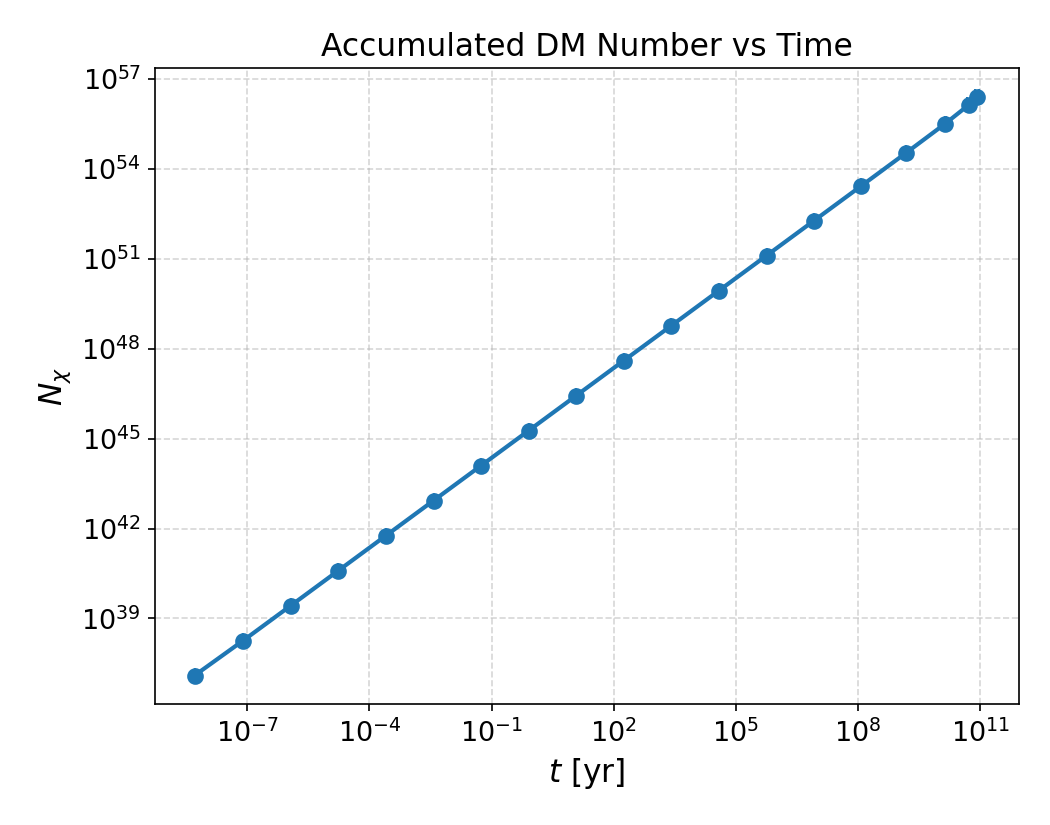}\\
    ~~~~(a)\\
    \includegraphics[width=0.8\linewidth]{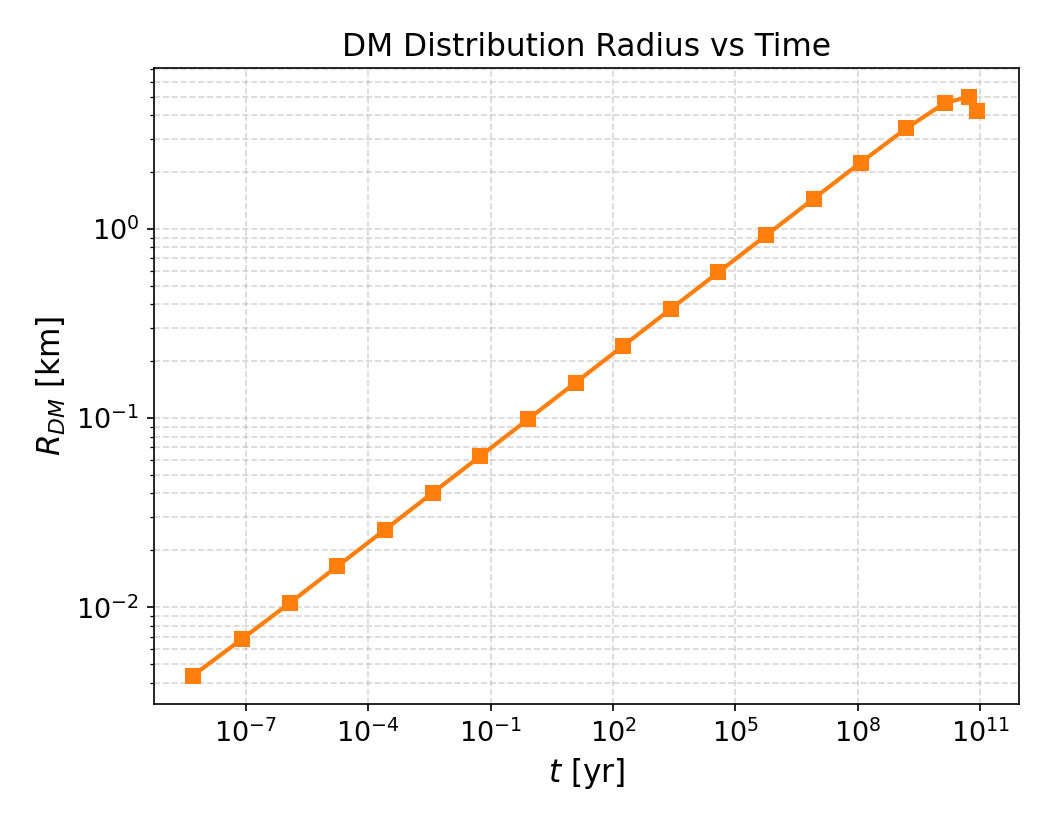}\\
    ~~~~(b)\\
    \includegraphics[width=0.8\linewidth]{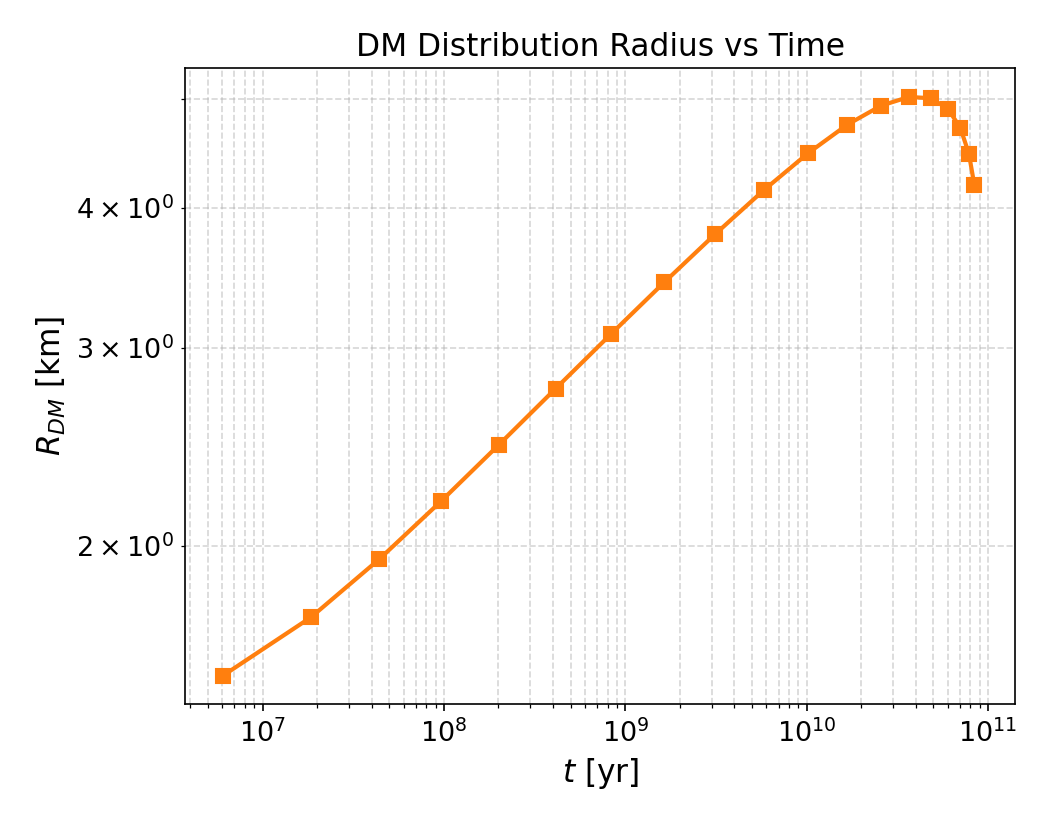}\\
    ~~~~(c)\\
    \caption{
Time evolution of the accumulated DM particle number \(N_\chi\) and the characteristic DM distribution radius \(R_{\rm DM}\) in the maximized capture scenario with \(\rho_{\rm halo}=2.2\times10^{9}\,M_\odot/\mathrm{pc}^3\). (a) Evolution of the accumulated DM particle number. (b) Evolution of the characteristic DM distribution radius. (c) Zoomed in view of (b), highlighting the late time decrease of \(R_{\rm DM}\).
}
    \label{figdensebg}
\end{figure}

The total accumulated DM mass may be estimated as
\begin{align}
    M_{\chi,\max}
    \sim
    m_\chi
    \int_0^{t_H}
    C_{\rm cap}(t)\,dt,
\end{align}
where \(t_H\) denotes the Hubble time. Under the maximized capture conditions described above, we estimate that the total NS mass increases only to
\begin{align}
    M_{\rm tot,max}
    \approx
    1.4484\,M_\odot
\end{align}
within a Hubble time. This quantity represents the maximal DM mass that can accumulate under highly favorable astrophysical conditions. Even in this optimized scenario, however, the accumulated DM component remains much smaller than the total NS mass,
\begin{align}
    M_\chi(t_H)\ll M_{\rm ns},
\end{align}
showing that the accumulated DM component remains gravitationally subdominant throughout the evolution. Consequently, although long term DM capture modifies the stellar compactness, radius, and tidal response, the overall structural changes remain modest throughout the parameter space explored in this work.

In both scenarios, the accumulated DM component produces a mild compression of the baryonic core and a corresponding reduction in the stellar radius, demonstrating that the gravitational response of the NS is robust against variations in the ambient DM environment. Although the compression becomes more pronounced in the maximized capture scenario, the overall structural modification remains modest even under these deliberately extreme assumptions.

\subsubsection{Degeneracy Threshold}\label{subsubsecdeg}

The two fluid TOV framework describes the accumulated DM component using a degenerate fermionic EoS. It is therefore important to verify that the captured DM indeed enters the degenerate regime under the astrophysical conditions considered here. We therefore estimate the density threshold at which the Fermi temperature exceeds the thermal temperature of the captured DM population \(T_F \gg T\).

For a non-relativistic fermion gas, the Fermi energy is
\begin{align}
    E_F
    =
    \frac{\hbar^2 k_F^2}{2m_\chi},
\end{align}
where \(m_\chi\) is the DM particle mass and \(k_F\) is the Fermi momentum. The corresponding Fermi temperature is therefore
\begin{align}
    T_F
    =
    \frac{E_F}{k_B}
    =
    \frac{\hbar^2 k_F^2}{2m_\chi k_B}.
\end{align}

The degeneracy condition \(T_F \gg T\) implies
\begin{align}
    k_F
    \gg
    \sqrt{\frac{2m_\chi k_B T}{\hbar^2}}.
\end{align}
Using the standard relation $k_F=(3\pi^2 n_\chi)^{1/3}$, the corresponding threshold number density becomes
\begin{align}
    n_\chi
    \gg
    \frac{1}{3\pi^2}
    \left(
    \frac{2m_\chi k_B T}{\hbar^2}
    \right)^{3/2}.
\end{align}

For the benchmark parameters adopted in this work, corresponding to a NS core temperature \(T\sim10^7~\mathrm{K}\) and a DM particle mass \(m_\chi\sim1~\mathrm{GeV}\), the corresponding degeneracy threshold is
\begin{align}
    n_{\rm deg}
    \approx
    9.9\times10^{30}~\mathrm{cm^{-3}}.
\end{align}

\begin{figure}
    \centering
    \includegraphics[width=0.8\linewidth]{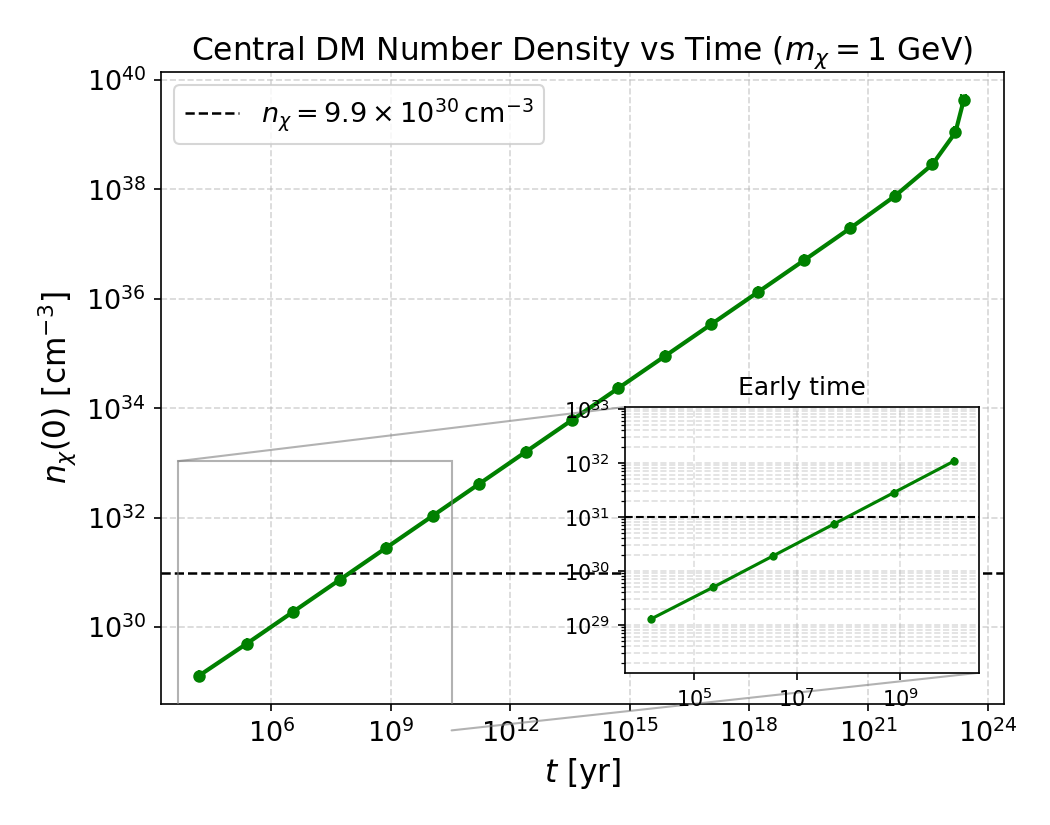}\\
    ~~~~(a)\\
    \includegraphics[width=0.8\linewidth]{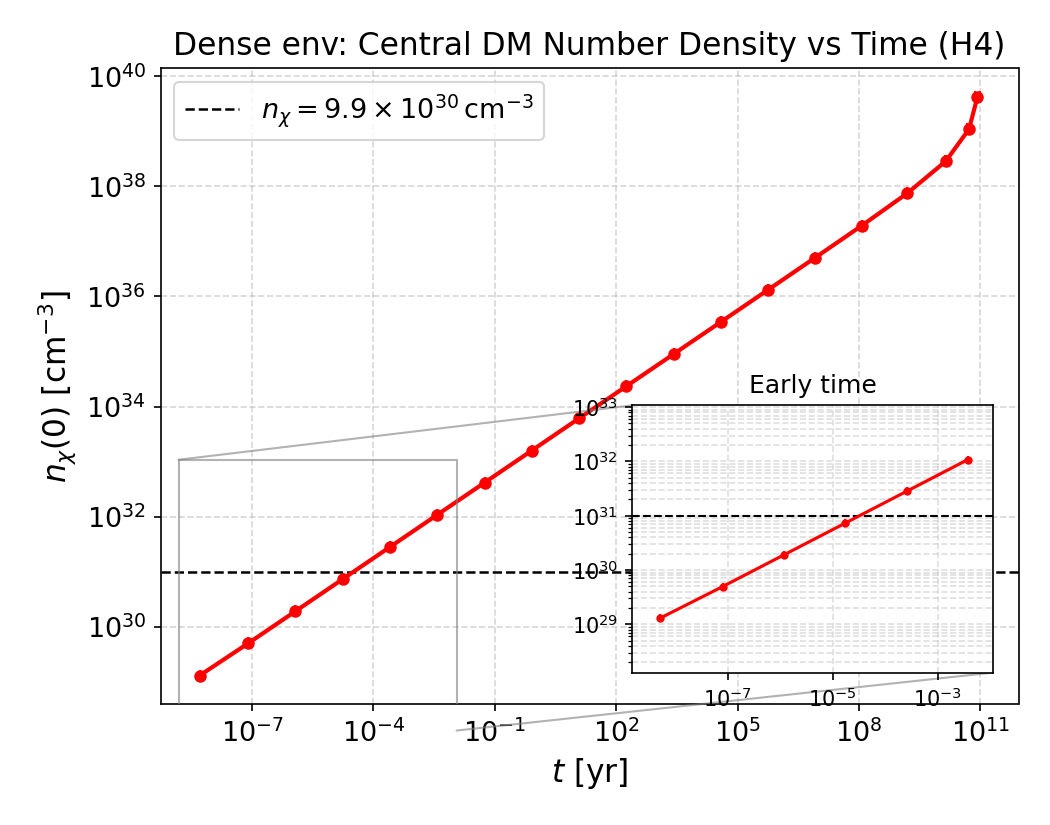}\\
    ~~~~(b)\\
    \caption{
Time evolution of the central DM number density \(n_\chi(0)\). (a) Canonical Galactic background environment with \(\rho_{\rm halo}=0.3~\mathrm{GeV\,cm^{-3}}\). (b) Maximized capture scenario with a dense DM environment. In both cases, the central DM number density rapidly exceeds the degeneracy threshold, validating the use of the degenerate fermionic EoS throughout the subsequent evolution.
}
    \label{figtndmc}
\end{figure}

Fig.~\ref{figtndmc} shows the time evolution of the central DM number density for the two benchmark scenarios considered in this work. In both cases, the central DM number density rapidly exceeds the degeneracy threshold, indicating that the captured DM enters the degenerate regime long before the NS undergoes significant astrophysical evolution. The degeneracy condition is therefore satisfied throughout the subsequent evolution, validating the use of the degenerate Fermi gas EoS in the two fluid TOV framework.

\subsection{Collapse Limit}

An important question is whether long term DM accumulation can eventually destabilize the NS and drive it toward gravitational collapse. Since the accumulated DM component increases both the gravitational mass and the stellar compactness, sufficiently large DM fractions could in principle push the system toward the instability threshold separating stable and collapsing NS configurations.

To examine this possibility, we study the stability of the equilibrium sequence under the maximized capture scenarios considered in this work~\cite{chandrasekhar1964dynamical,Goldman:1989nd,bramante2014detecting,bramante2015dark,Bramante:2017ulk,Robles:2025dlv}. As we show below, even under these deliberately extreme assumptions, the NS remains far from the collapse limit.

The stability of the equilibrium sequence can be analyzed using the standard turning point criterion
\begin{align}
    \frac{dM_{\rm tot}}{d\rho_c} > 0,
\end{align}
where \(M_{\rm tot}\) is the total stellar mass and \(\rho_c\) is the central baryon density. The onset of instability occurs when
\begin{align}
    \frac{dM_{\rm tot}}{d\rho_c}=0,
\end{align}
which marks the maximum mass configuration separating stable and unstable branches of the equilibrium sequence.

\begin{figure}
    \centering
    \includegraphics[width=0.8\linewidth]{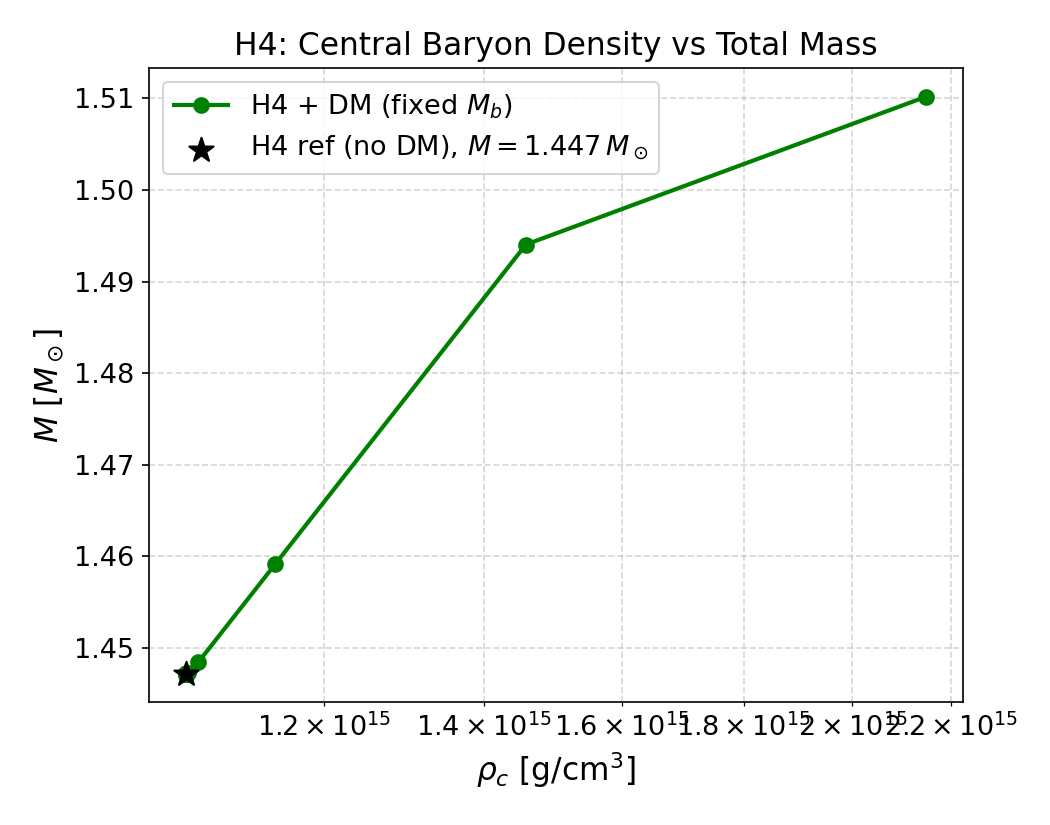}
    \caption{
Total NS mass \(M_{\rm tot}\) as a function of the central baryon density \(\rho_c\) for different accumulated DM contents. The turning point of each equilibrium sequence corresponds to the onset of gravitational instability. For all DM configurations considered in this work, the equilibrium sequences remain on the stable branch satisfying \(dM_{\rm tot}/d\rho_c>0\).
}
    \label{figH4rhocM}
\end{figure}

To assess how DM accumulation affects the stability of the equilibrium sequence, we solve the coupled two fluid TOV equations for different accumulated DM contents and construct the corresponding mass--central density relations. The resulting \(M_{\rm tot}\)-\(\rho_c\) sequences are shown in Fig.~\ref{figH4rhocM}.

As shown in Fig.~\ref{figH4rhocM}, all equilibrium configurations reached over the astrophysical timescales considered in this work remain on the stable branch of the equilibrium sequence satisfying \(dM_{\rm tot}/d\rho_c>0\). Although the accumulated DM component slightly modifies the stellar structure and increases the compactness, it remains insufficient to push the NS toward the maximum mass configuration or trigger gravitational collapse, even in the maximized capture scenario considered in this work.

\section{Neutron Star Properties}\label{secnsproperty}

We now examine how the accumulated DM component modifies key structural properties of NSs and the resulting gravitational wave observables.

\subsection{Compactness}
The stellar compactness is defined as
\begin{align}
    C=\frac{M_{\rm tot}}{R_{\rm ns}},
\end{align}
which measures the strength of the gravitational field at the stellar surface and directly affects observable quantities such as the tidal deformability, Love number, and gravitational wave response of the star.

\begin{figure}[htp!]
    \centering
    \includegraphics[width=0.8\linewidth]{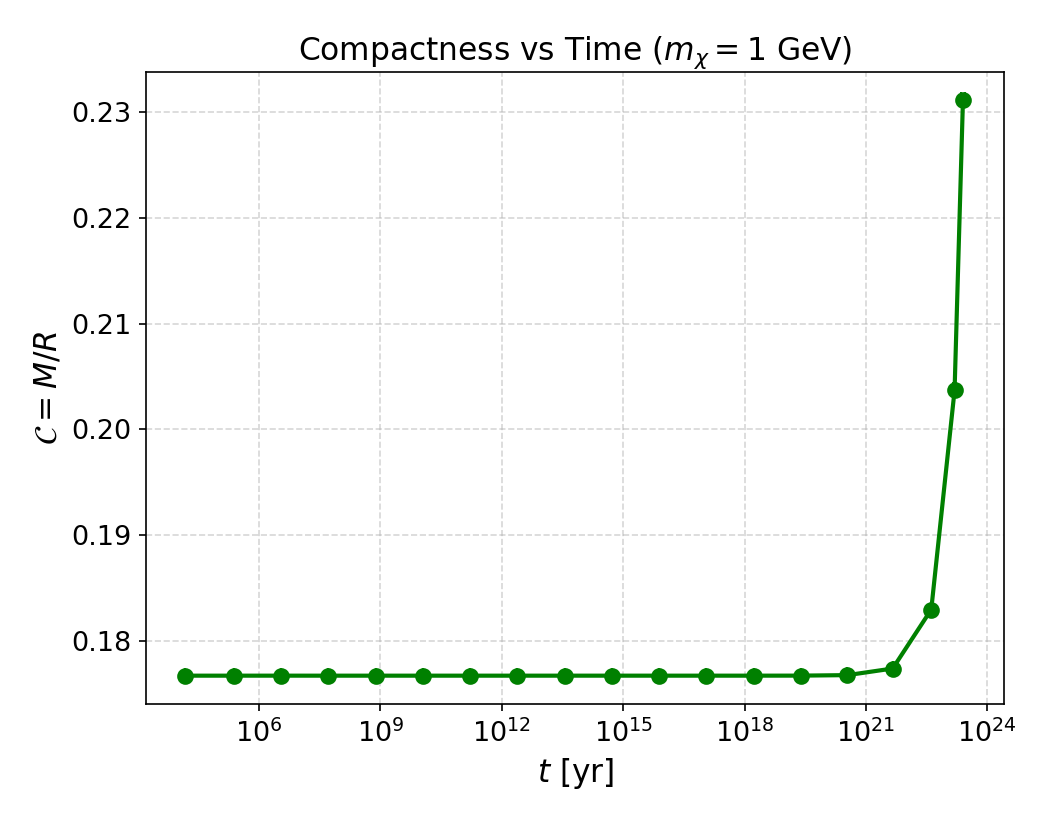}\\
    ~~~~(a)\\
    \includegraphics[width=0.8\linewidth]{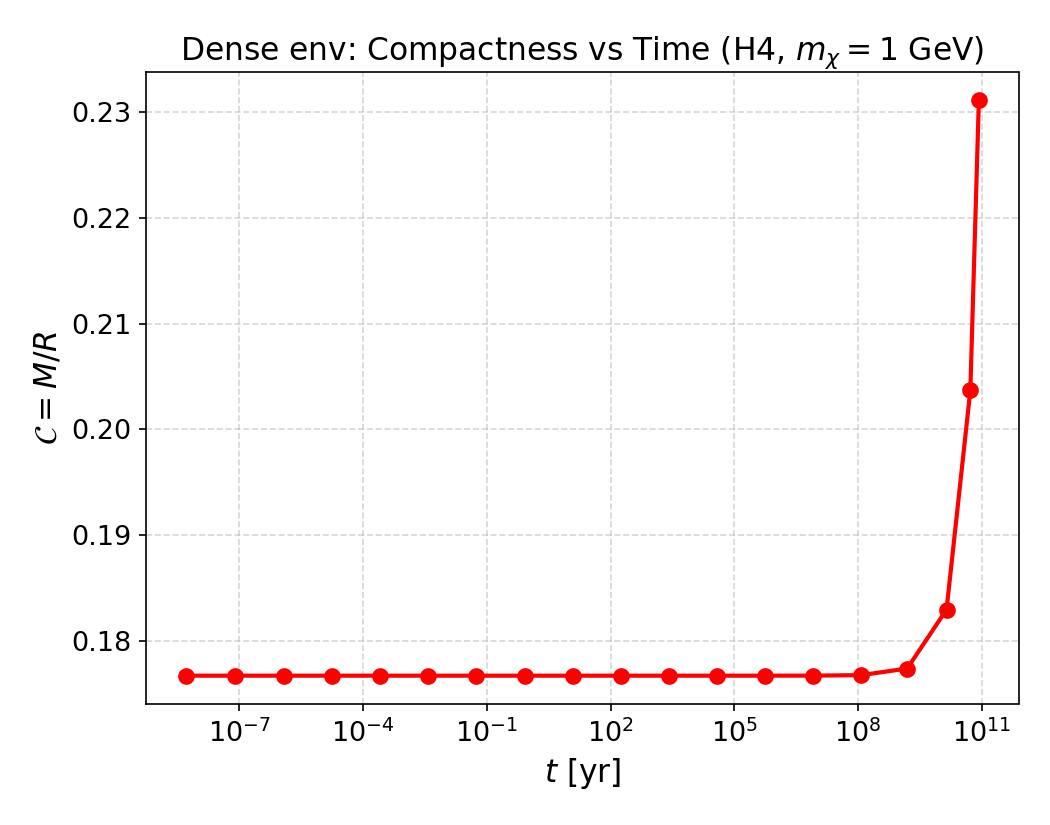}\\
    ~~~~(b)\\
    \caption{
Reconstructed time evolution of the NS compactness \(C=M_{\rm tot}/R_{\rm ns}\) under DM accumulation. (a) Canonical Galactic background environment. (b) Maximized capture scenario with dense DM spikes. In both cases, the stellar compactness increases only modestly over a Hubble time.
}
    \label{figtC}
\end{figure}

The reconstructed time evolution of the compactness is shown in Fig.~\ref{figtC}. As DM accumulates inside the NS, the additional gravitational contribution mildly compresses the baryonic core, reducing the stellar radius and increasing the compactness. Although this effect becomes more pronounced in the maximized capture scenario, the overall increase in compactness remains modest even over a Hubble time, reflecting the gravitationally subdominant nature of the accumulated DM component.

%\begin{figure}
%    \centering
%    \includegraphics[width=1\linewidth]{t_to_C.png}
%    \caption{Compactness evolve with time. The blue line shows the DM admixed NS, the yellow line shows an NS without the DM component. The parameter choice here is $\rho_{\chi}=2.2\times10^{9}\,M_{\odot}\,\mathrm{pc^{-3}},~v_{ns}=30km/s,~m_{\chi}=1.48\mathrm{GeV}$}
%    \label{figttoC}
%\end{figure}
Even under the deliberately optimized capture scenario considered here—corresponding to light asymmetric fermionic DM embedded in a dense DM spike environment over a Hubble time—the NS compactness increases only at the tens of percent level, typically by about \(25\%-30\%\). Although this represents a noticeable structural modification, such behavior appears only in intentionally extreme capture environments designed to maximize the accumulated DM population.

In more realistic astrophysical environments, where the ambient DM density is closer to canonical Galactic or cluster halo values, the accumulated DM fraction is substantially smaller. The corresponding increase in the stellar compactness is therefore significantly reduced, remaining well below the level reached in the maximized capture scenario.

\subsection{Love Number and Tidal Deformability}\label{subseck2}

To characterize the observational consequences of DM accumulation, we compute the tidal response of NSs in compact binary systems. During the inspiral stage of a merger, the external gravitational field of the companion induces tidal deformations that depend sensitively on the internal structure, compactness, and EoS of the NS. These tidal effects leave measurable imprints on the gravitational wave phase evolution and therefore provide one of the most important observational probes of NS structure.

The tidal response is characterized by the dimensionless Love number \(k_2\) and the associated tidal deformability parameter \(\Lambda\). Since DM accumulation modifies the stellar compactness and density profile, it can in principle alter both \(k_2\) and \(\Lambda\), providing a direct connection between long term DM accumulation and gravitational wave observables.

For a compact binary system, the effective tidal deformability \(\tilde{\Lambda}\) is given by
\begin{align}
    \tilde{\Lambda}
    =
    \frac{16}{13}
    \frac{
    (M_{1}+12M_{2})M_{1}^4\Lambda_{1}
    +
    (M_{2}+12M_{1})M_{2}^4\Lambda_{2}
    }{
    (M_{1}+M_{2})^5
    },
\end{align}
where, $\Lambda_i=\frac{2}{3}\frac{k_2^i}{C_i^5}$ is the dimensionless tidal deformability of an individual compact object with compactness
\begin{align}
    C_i=\frac{M_i}{R_i}.
\end{align}

For a black hole, \(k_2^{\rm BH}=0\) implies \(\Lambda_{\rm BH}=0\); therefore, in neutron star black hole (NS–BH) binaries only the NS component contributes to the tidal response. In this case, the effective tidal deformability reduces to
\begin{align}
    \tilde{\Lambda}_{\rm NS}
    =
    \frac{32}{39}
    \frac{
    (M_{\rm NS}+12M_{\rm BH})M_{\rm NS}^4
    }{
    (M_{\rm NS}+M_{\rm BH})^5
    }
    \frac{k_2}{C_{\rm NS}^5}.
\end{align}

The Love number \(k_2\) is obtained by solving for the function \(y(r)\), which encodes the quadrupolar tidal response of the NS~\cite{hinderer2008tidal,Leung:2022wcf}. In each baryonic layer, \(y(r)\) satisfies
\begin{align}\label{eqypure}
    r y'
    +
    y^2
    +
    y e^{\lambda}\!\left[
    1+4\pi r^2(p-\epsilon)
    \right]
    +
    r^2 Q
    =
    0,
\end{align}
where primes denote derivatives with respect to \(r\), \(\epsilon\) is the energy density, and \(e^\lambda=B(r)\) and \(\nu=2\Phi(r)\). The function \(Q\) is given by
\begin{align}
    Q
    =
    4\pi e^{\lambda}\!\left(
    5\epsilon+9p+\frac{\epsilon+p}{dp/d\epsilon}
    \right)
    -
    \frac{6e^{\lambda}}{r^2}
    -
    (\nu')^2 .
\end{align}

In the presence of a DM component, the tidal perturbation is sourced by the total stress energy tensor of all fluid components. Following Refs.~\cite{Leung:2022wcf}, the perturbation equation generalizes to
\begin{align}\label{eqytot}
    r y'
    +
    y^2
    +
    y e^{\lambda}
    \sum_i\!\left[
    1+4\pi r^2(p_i-\epsilon_i)
    \right]
    +
    r^2 Q
    =
    0,
\end{align}
with
\begin{align}
    Q
    =
    4\pi e^{\lambda}
    \sum_i\!\left(
    5\epsilon_i+9p_i+\frac{\epsilon_i+p_i}{dp_i/d\epsilon_i}
    \right)
    -
    \frac{6e^{\lambda}}{r^2}
    -
    (\nu')^2 ,
\end{align}
where the sum runs over the baryonic and DM components.

At the boundary between adjacent layers, we impose the matching conditions
\begin{align}\label{eqymatch}
    y_i(R_i)
    =
    y_{i+1}(R_i),
    \qquad
    y_i'(R_i)
    =
    y_{i+1}'(R_i).
\end{align}

\begin{figure}[htp!]
    \centering
    \includegraphics[width=0.8\linewidth]{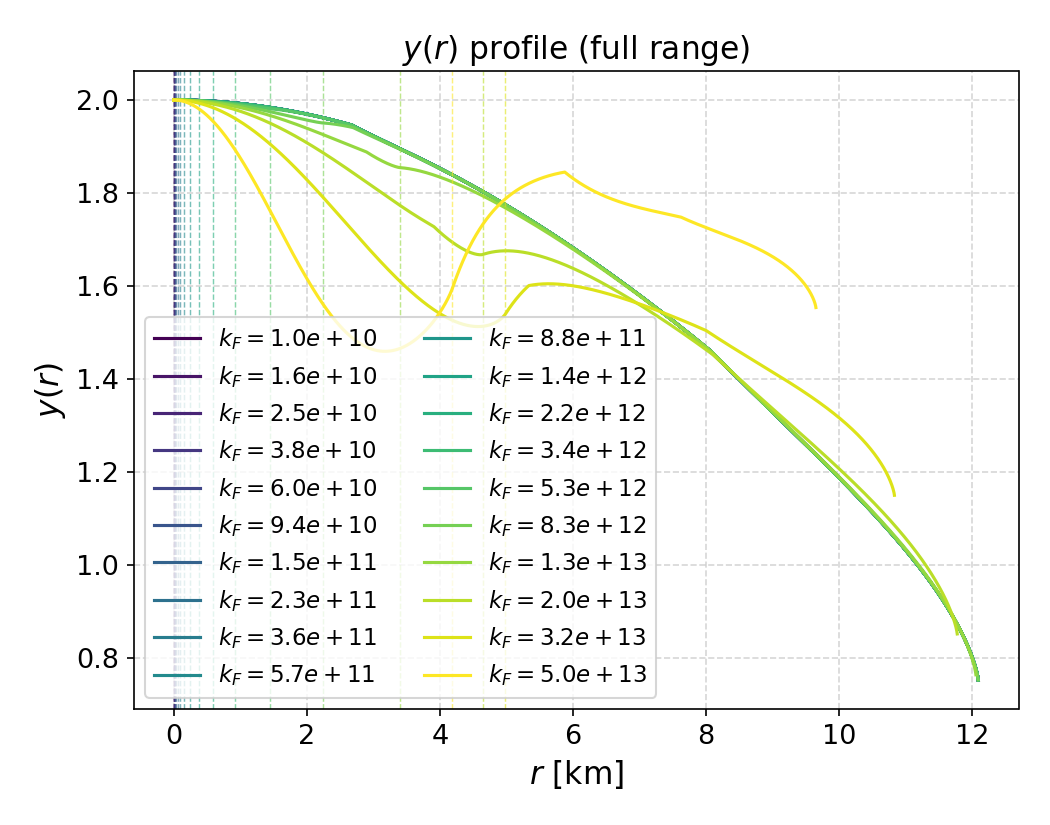}
    \caption{
Radial profile of the tidal perturbation function \(y(r)\) for the reference NS model. The surface value \(y_R=y(R_{\rm ns})\) obtained from this profile is used to compute the Love number \(k_2\) and the tidal deformability \(\Lambda\).
}
    \label{figy}
\end{figure}

Equation~\eqref{eqytot} is integrated outward from the stellar center using a first order Euler scheme together with the boundary and matching conditions above, yielding the surface value
\begin{align}
    y_R
    =
    y(R_{\rm ns}),
\end{align}
the radial profile of which is shown in Fig.~\ref{figy}.

The radial evolution of \(y(r)\) illustrates how DM accumulation modifies the internal tidal response of the NS. Although the accumulated DM component remains gravitationally subdominant, its contribution to the spacetime metric and density profile leads to corresponding changes in the Love number \(k_2\) and the tidal deformability \(\Lambda\).

From the solution for \(y(r)\), the Love number \(k_2\) is given by~\cite{hinderer2008tidal}
\begin{align}
    k_2
    &=
    \frac{8C_{\rm NS}^5}{5}
    (1-2C_{\rm NS})^2
    \left[
    2+2C_{\rm NS}(y_R-1)-y_R
    \right]\nonumber\\&\times
    \left\{
    2C_{\rm NS}
    \left[
    6-3y_R+3C_{\rm NS}(5y_R-8)
    \right]
    \right.
    \nonumber\\
    &
    \left.
    +4C_{\rm NS}^3
    \left[
    13-11y_R+C_{\rm NS}(3y_R-2)
    +2C_{\rm NS}^2(1+y_R)
    \right]\right.\nonumber\\&\left.
    +3(1-2C_{\rm NS})
    \left[
    2-y_R+2C_{\rm NS}(y_R-1)
    \right]
    \log(1-2C_{\rm NS})
    \right\}^{-1}.
\end{align}

\begin{figure}[htp!]
    \centering
    \includegraphics[width=0.8\linewidth]{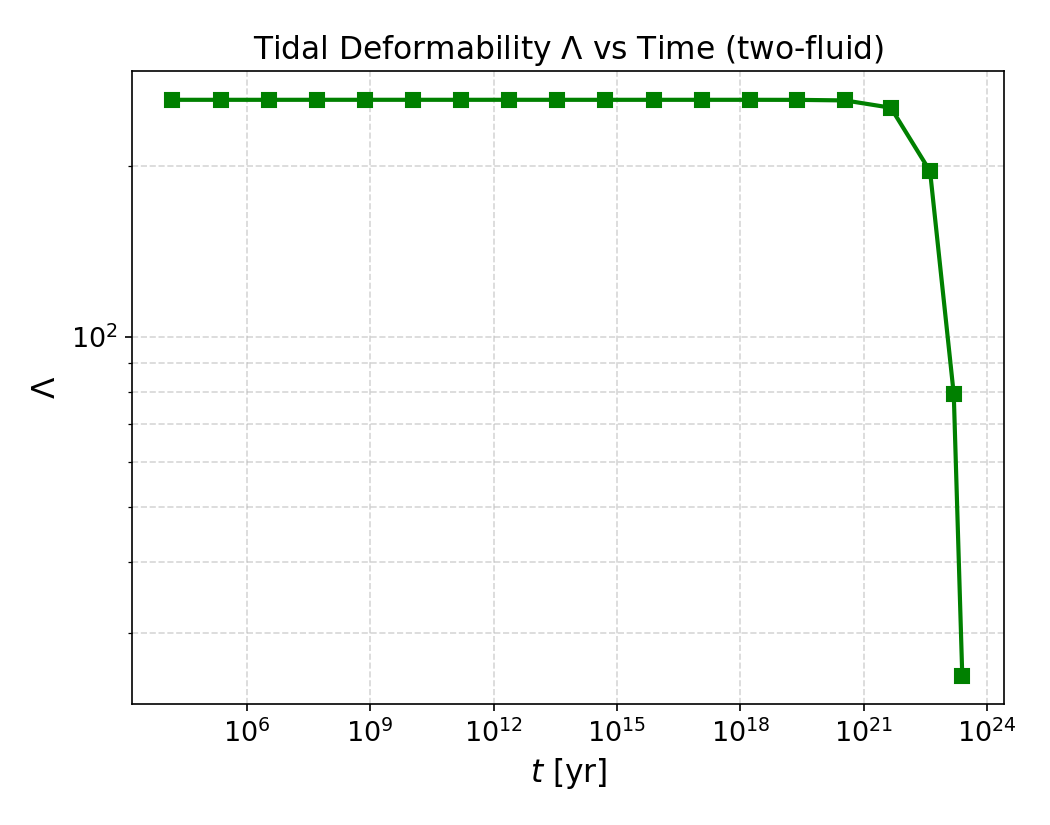}\\
    ~~~~(a)\\
    \includegraphics[width=0.8\linewidth]{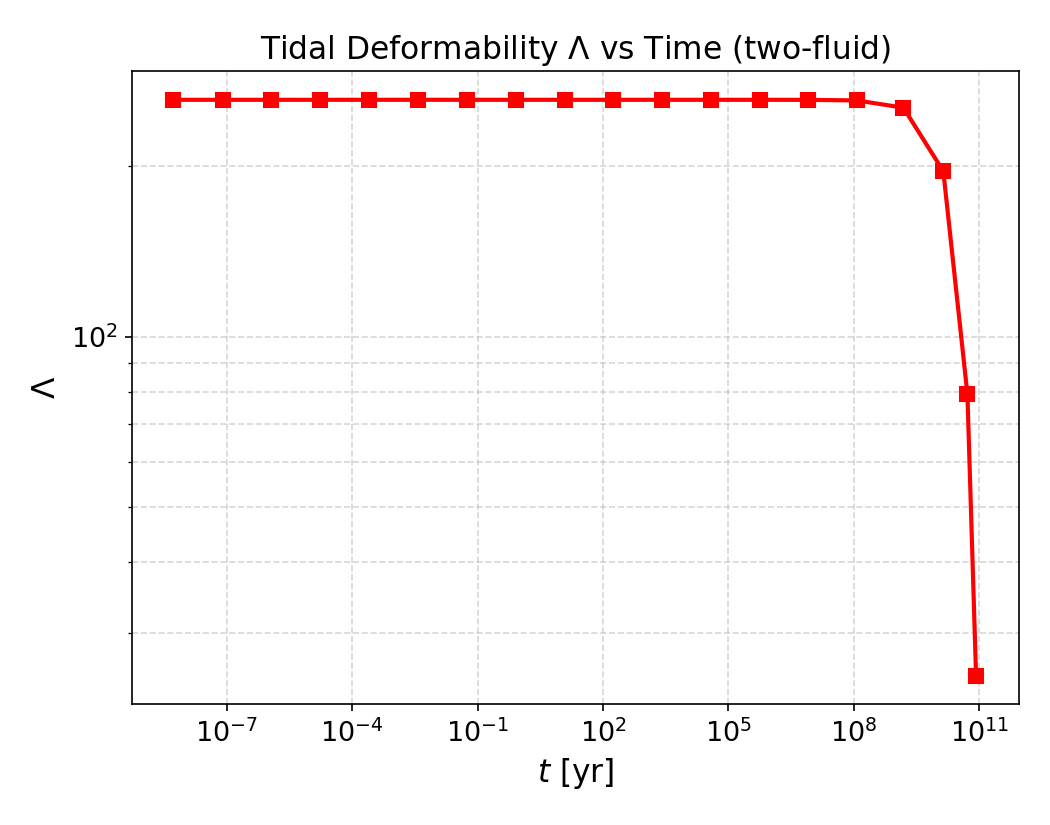}\\
    ~~~~(b)\\
    \caption{
Time evolution of the tidal deformability \(\Lambda\) under DM accumulation. (a) Canonical Galactic background environment. (b) Maximized capture scenario. In both cases, \(\Lambda\) decreases as DM accumulates, with substantially larger changes in the maximized capture scenario.
}
    \label{figttoLam}
\end{figure}

\begin{figure}[htp!]
    \centering
    \includegraphics[width=0.8\linewidth]{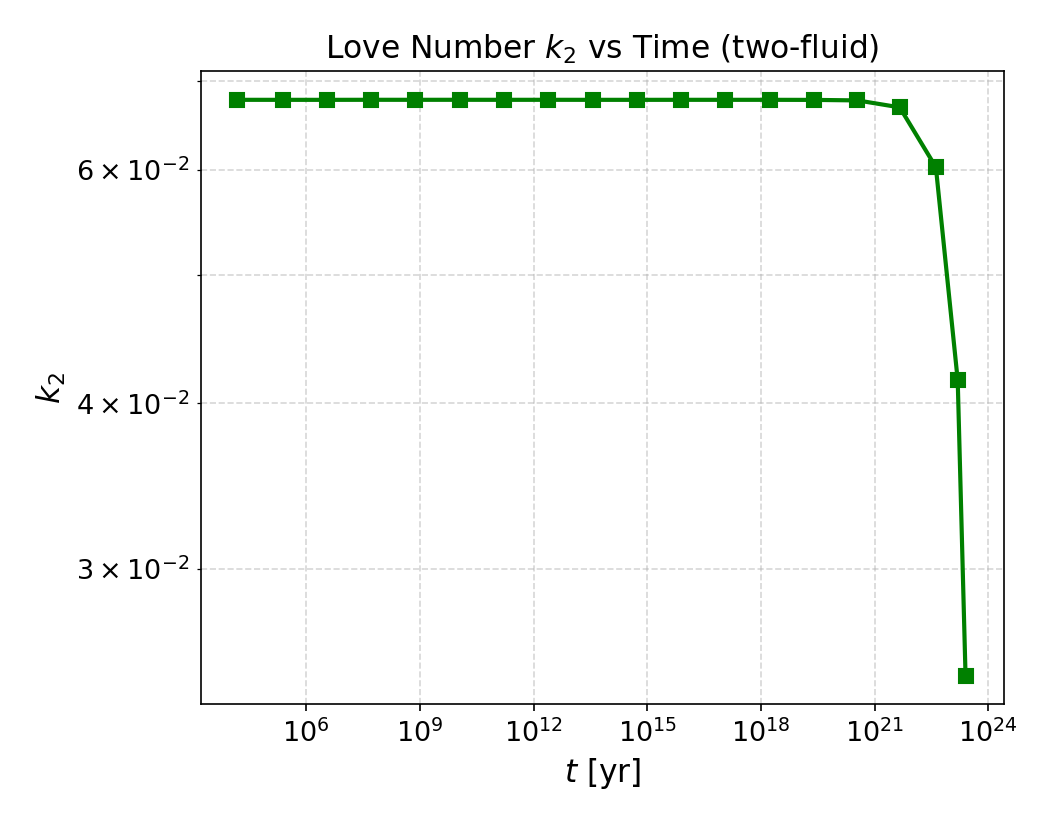}\\
    ~~~~(a)\\
    \includegraphics[width=0.8\linewidth]{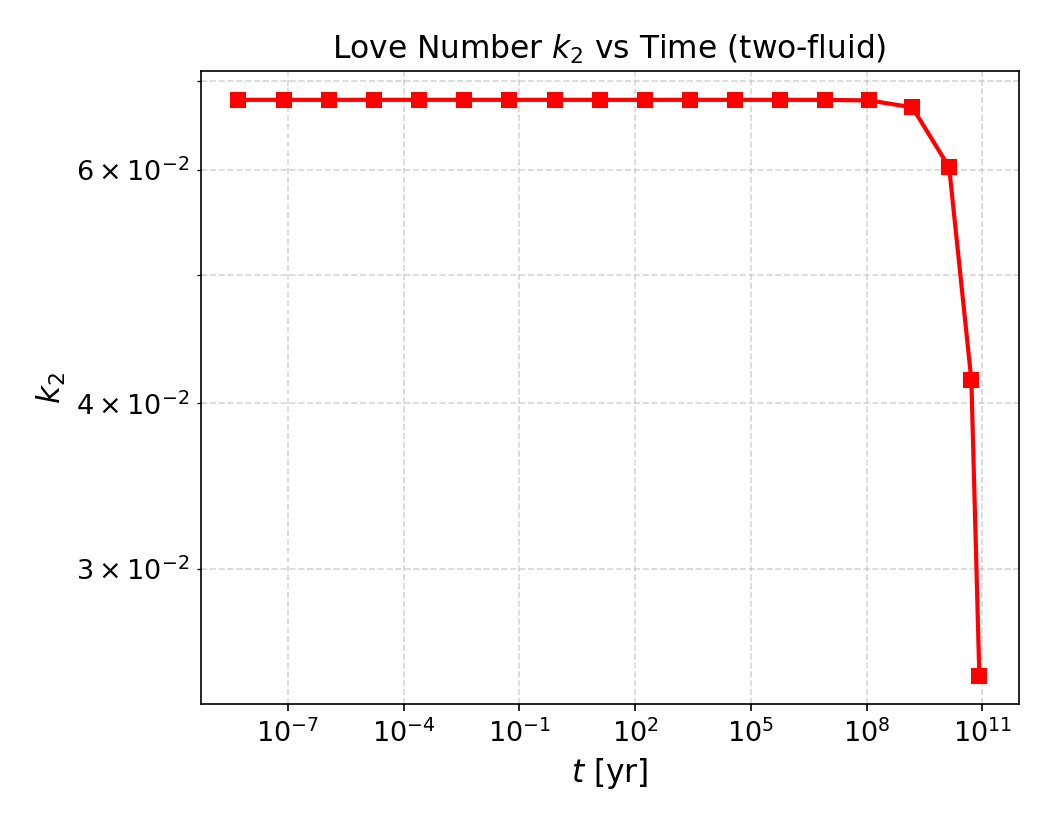}\\
    ~~~~(b)\\
    \caption{
Time evolution of the Love number \(k_2\) under DM accumulation. (a) Canonical Galactic background environment. (b) Maximized capture scenario. In both cases, \(k_2\) decreases as DM accumulates, with substantially larger changes in the maximized capture scenario.
}
    \label{figttok2}
\end{figure}

To quantify the impact of DM accumulation on the NS tidal response, we reconstruct the time evolution of the Love number \(k_2\) and the tidal deformability \(\Lambda\) for both the canonical Galactic background environment and the maximized capture scenario introduced above. The resulting evolution of \(k_2\) and \(\Lambda\) is shown in Figs.~\ref{figttok2} and \ref{figttoLam}.

As shown in Figs.~\ref{figttok2} and \ref{figttoLam}, both the Love number \(k_2\) and the tidal deformability \(\Lambda\) decrease monotonically with time as DM accumulates inside the NS. This behavior reflects the gradual increase in the stellar compactness, as the gravitational potential deepens, the NS becomes more tightly bound and less susceptible to external tidal fields, suppressing its quadrupolar tidal response.

Since the tidal deformability scales as
\begin{align}
    \Lambda \propto k_2 C^{-5},
\end{align}
even modest increases in the stellar compactness produce comparatively larger changes in \(\Lambda\) than in the Love number \(k_2\). Significant evolution occurs only in the maximized capture scenario corresponding to an extreme Galactic center DM environment, where
\begin{align}
    |\Delta k_2|
    \sim
    10^{-2},
    \qquad
    |\Delta\Lambda|
    \sim
    10^{1},
\end{align}
over a Hubble time. For canonical Galactic background DM densities, the corresponding changes remain negligible. These results indicate that, even under deliberately optimized capture conditions, the effects of long term DM accumulation on tidal observables remain too small to be distinguished with current gravitational wave observations.

\subsection{Dependence on the Equation of State}\label{subsecdependEOS}

The structural response of a NS to DM accumulation depends not only on the properties of the dark sector, but also on the underlying nuclear EoS. In particular, softer and stiffer EoS models can respond differently to the additional gravitational contribution from the accumulated DM component. To investigate how sensitive the resulting structural modifications are to the NS EoS, we perform a comparative analysis using several representative EoS models, including SLy, AP4, MPA1, and H4, which span a broad range of stiffness.

The EoS parameters are adopted from Ref.~\cite{read2009constraints}, where each model is anchored by the pressure \(p_1\) at the transition density
\begin{align}
    \rho_1 = 10^{14.7}\,\mathrm{g\,cm^{-3}},
\end{align}
from which the outer polytropic normalization is determined through
\begin{align}
    K_1 = \frac{p_1}{\rho_1^{\Gamma_1}}.
\end{align}

This normalization differs from that adopted for the single H4 reference model used in the previous sections, where the outer polytropic segment was anchored at the effective surface density. As a result, the H4 masses and radii reported here differ quantitatively from the earlier reference values. This choice ensures that all EoS models are normalized consistently, enabling a uniform comparison of their responses to DM accumulation.

\par
\begin{table}[htp!]
    \centering
    \scalebox{1.0}{
    \begin{tabular}{l|c|c}
    \hline\hline
    EoS & $M_{\rm ns}/M_{\odot}$ & $R_{\rm ns}/\mathrm{km}$\\
    \hline\hline
    SLy & $1.669$ & $11.177$\\
    \hline
    AP4 & $1.675$ & $10.782$\\
    \hline
    MPA1 & $2.338$ & $12.064$\\
    \hline
    H4 & $1.951$ & $13.438$\\
    \hline\hline
    \end{tabular}}
    \caption{
Representative NS configurations obtained for the EoS models adopted from Ref.~\cite{read2009constraints}. All models are constructed using the same central baryon density, \(\rho_c=1.05\times10^{15}\,\mathrm{g\,cm^{-3}}\), resulting in different stellar masses and radii for different EoSs.
}
    \label{tabEOScompare}
\end{table}

For each EoS, we repeat the full DM accumulation analysis under two representative astrophysical environments:
(i) a canonical Galactic background environment, and
(ii) an extreme high density environment corresponding to DM spikes.

\begin{widetext}
    \begin{figure*}[htp!]
    \centering
    \includegraphics[width=0.33\linewidth]{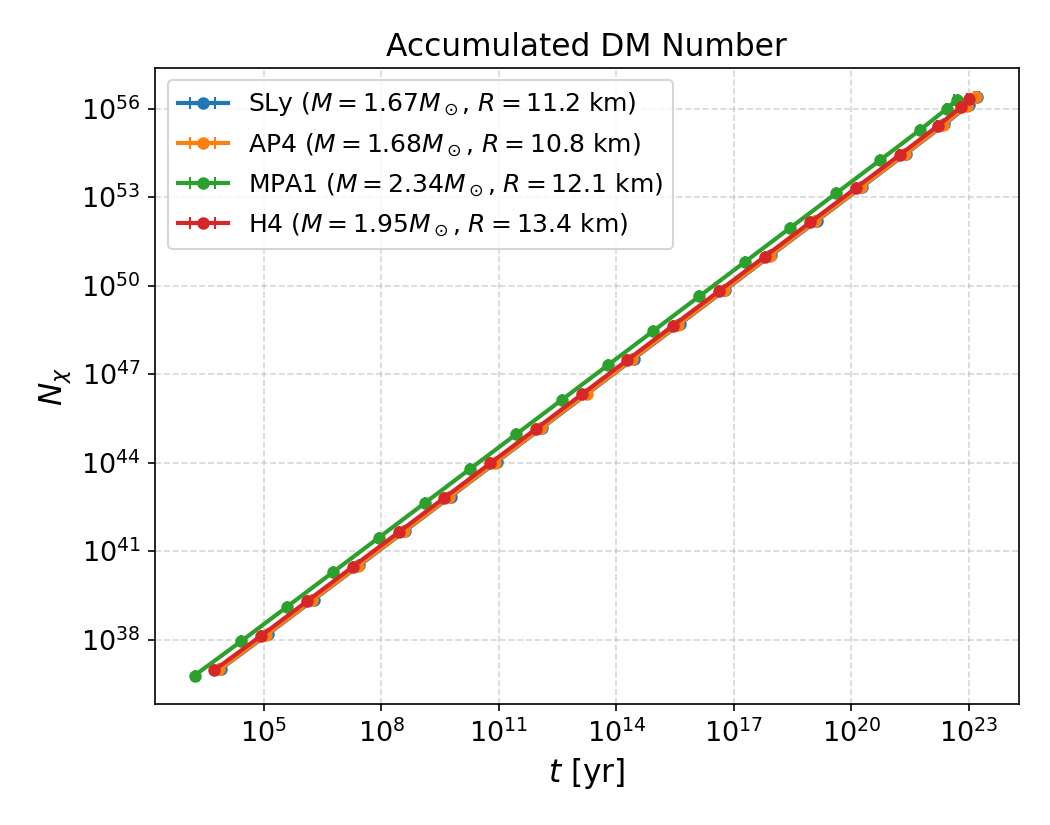}\includegraphics[width=0.33\linewidth]{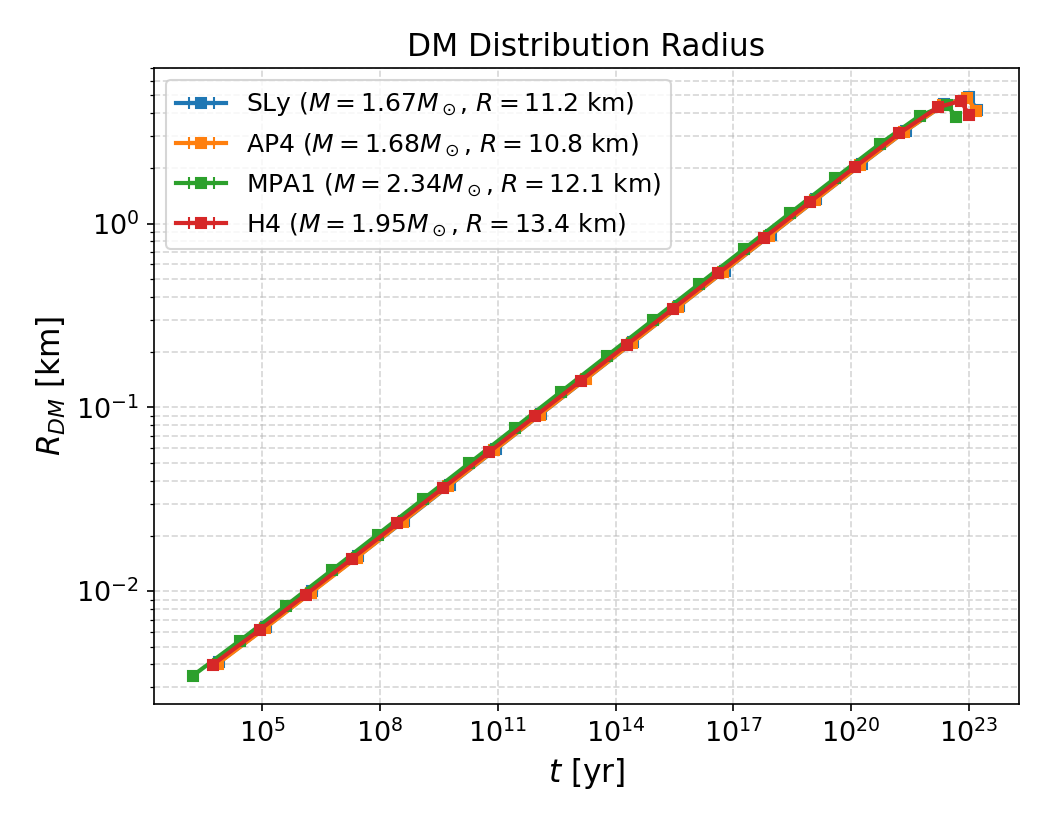}\includegraphics[width=0.33\linewidth]{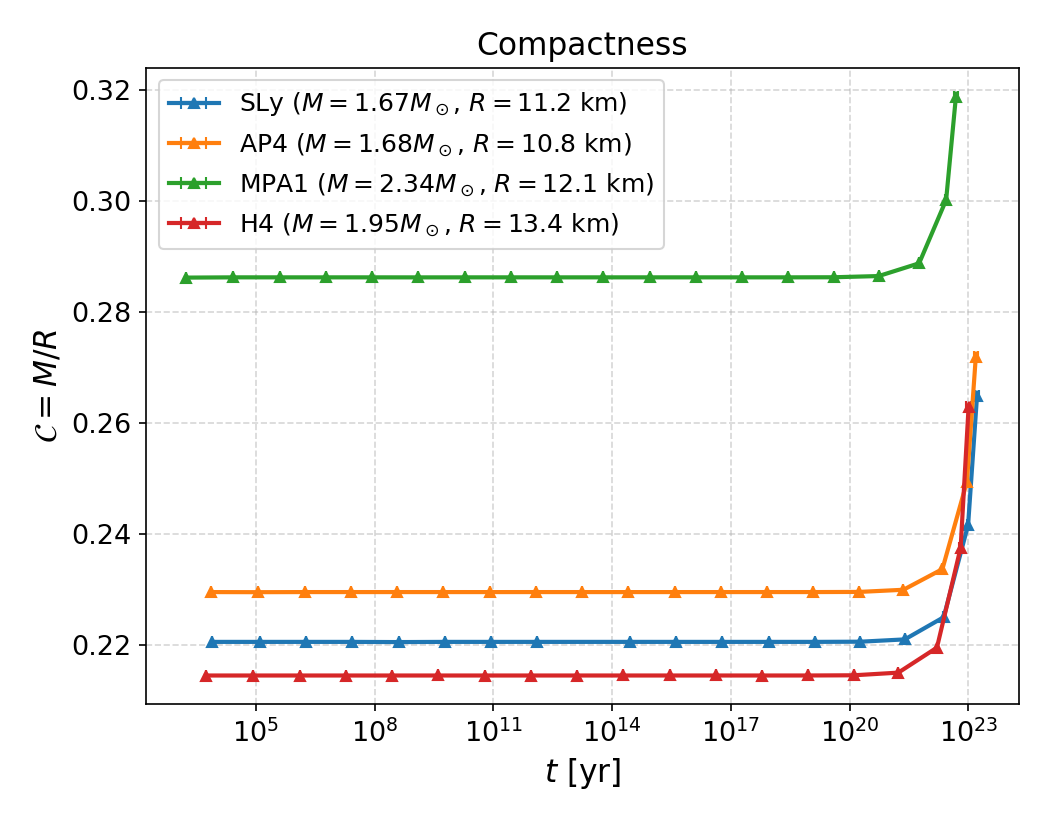}\\
    (a)~~~~~~~~~~~~~~~~~~~~~~~~~~~~~~~~~~~~~~~~~~~~~~~~(b)~~~~~~~~~~~~~~~~~~~~~~~~~~~~~~~~~~~~~~~~~~~~~~~~(c)\\
    \includegraphics[width=0.33\linewidth]{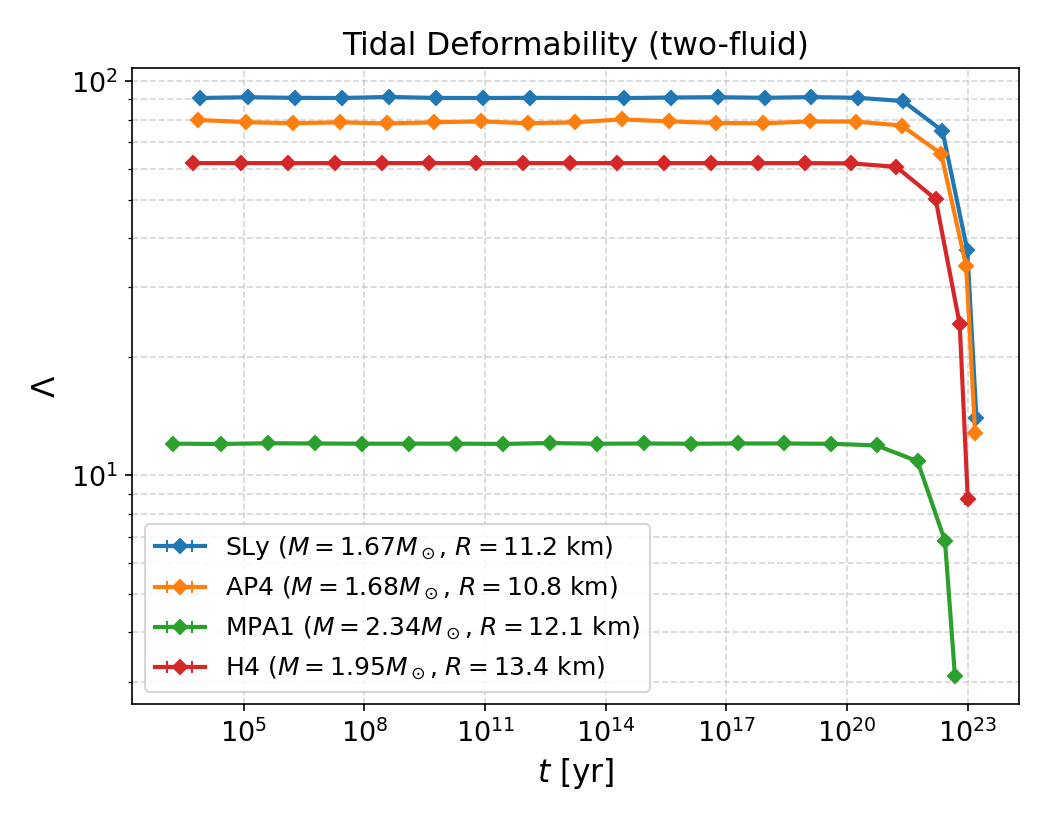}\includegraphics[width=0.33\linewidth]{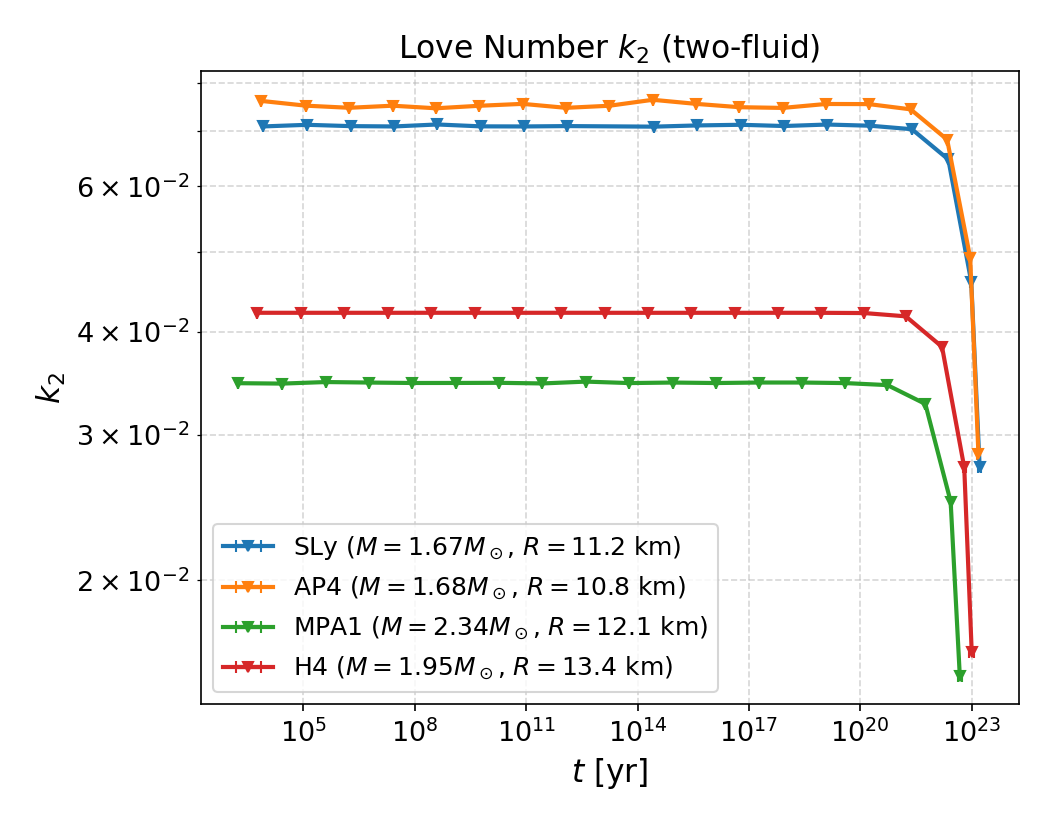}\includegraphics[width=0.33\linewidth]{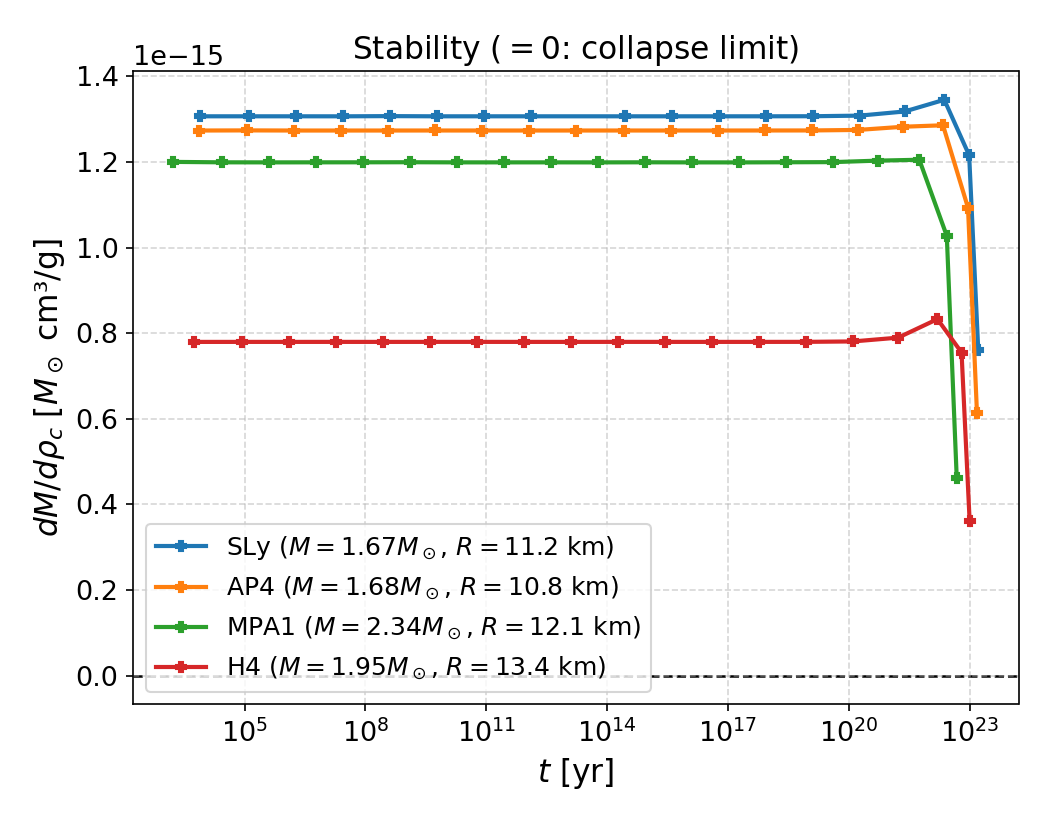}\\
    (d)~~~~~~~~~~~~~~~~~~~~~~~~~~~~~~~~~~~~~~~~~~~~~~~~(e)~~~~~~~~~~~~~~~~~~~~~~~~~~~~~~~~~~~~~~~~~~~~~~~~(f)\\
    \caption{
Comparison of the DM induced structural evolution for different NS EoS models (SLy, AP4, MPA1, and H4) under the canonical Galactic background DM environment. The time evolution of (a) the accumulated DM particle number \(N_\chi\), (b) the DM distribution radius \(R_{\rm DM}\), (c) the stellar compactness \(C\), (d) the tidal deformability \(\Lambda\), (e) the Love number \(k_2\), and (f) the stability indicator \(dM_{\rm tot}/d\rho_c\). Although the quantitative responses vary among the EoS models, the overall evolution remains qualitatively similar.
}
    \label{figEOS}
\end{figure*}
\end{widetext}

The results are shown in Fig.~\ref{figEOS}. Across all EoS models, DM accumulation produces the same qualitative behavior. The additional gravitational contribution mildly compresses the baryonic core, leading to a modest reduction in the stellar radius and a corresponding modification of the tidal response.

The magnitude of this effect depends on the stiffness of the EoS. Softer EoS models, which produce more compact NSs, exhibit slightly larger structural modifications, whereas stiffer EoS models respond more weakly. Nevertheless, the overall differences among the EoS models remain modest, indicating that our main conclusions are robust against uncertainties in the nuclear EoS.

\begin{widetext}
    \begin{figure*}[htp!]
    \centering
    \includegraphics[width=0.33\linewidth]{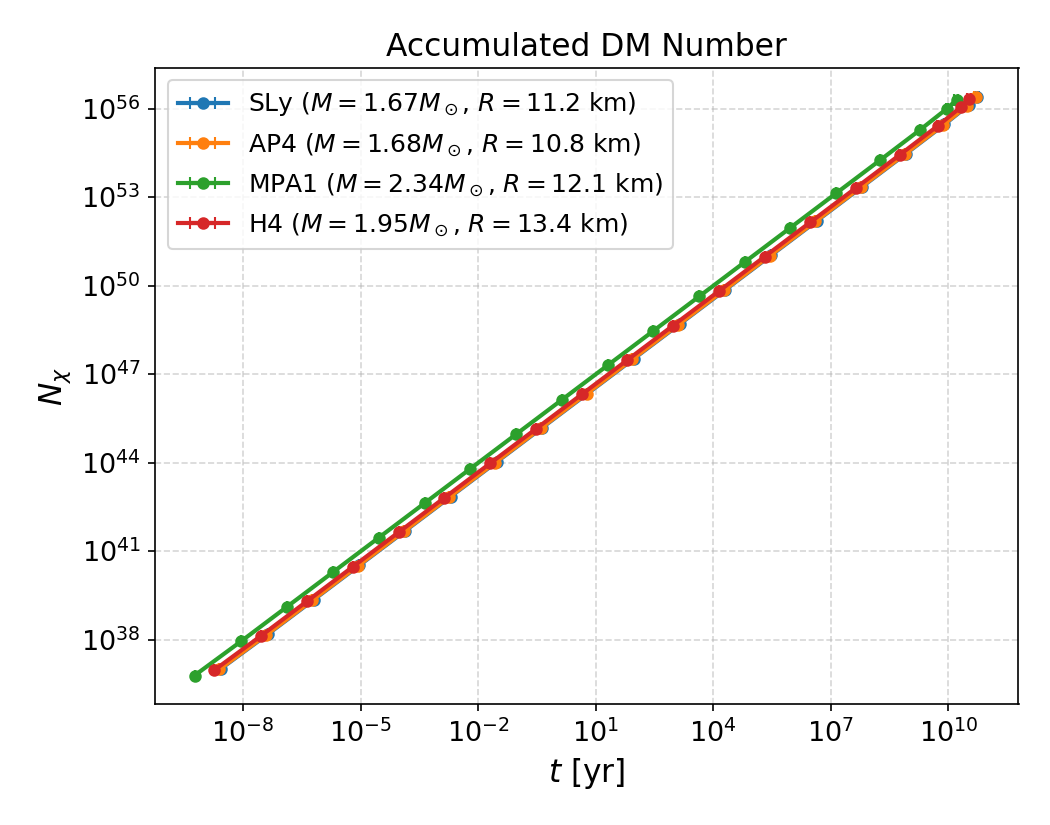}\includegraphics[width=0.33\linewidth]{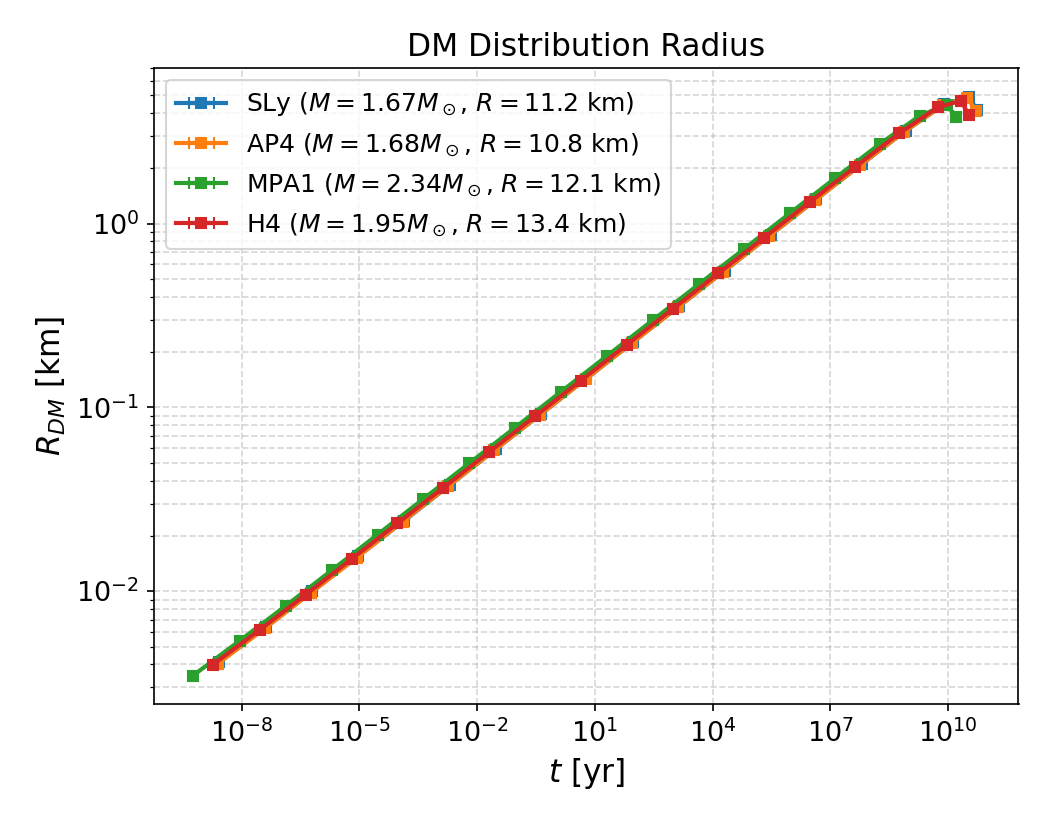}\includegraphics[width=0.33\linewidth]{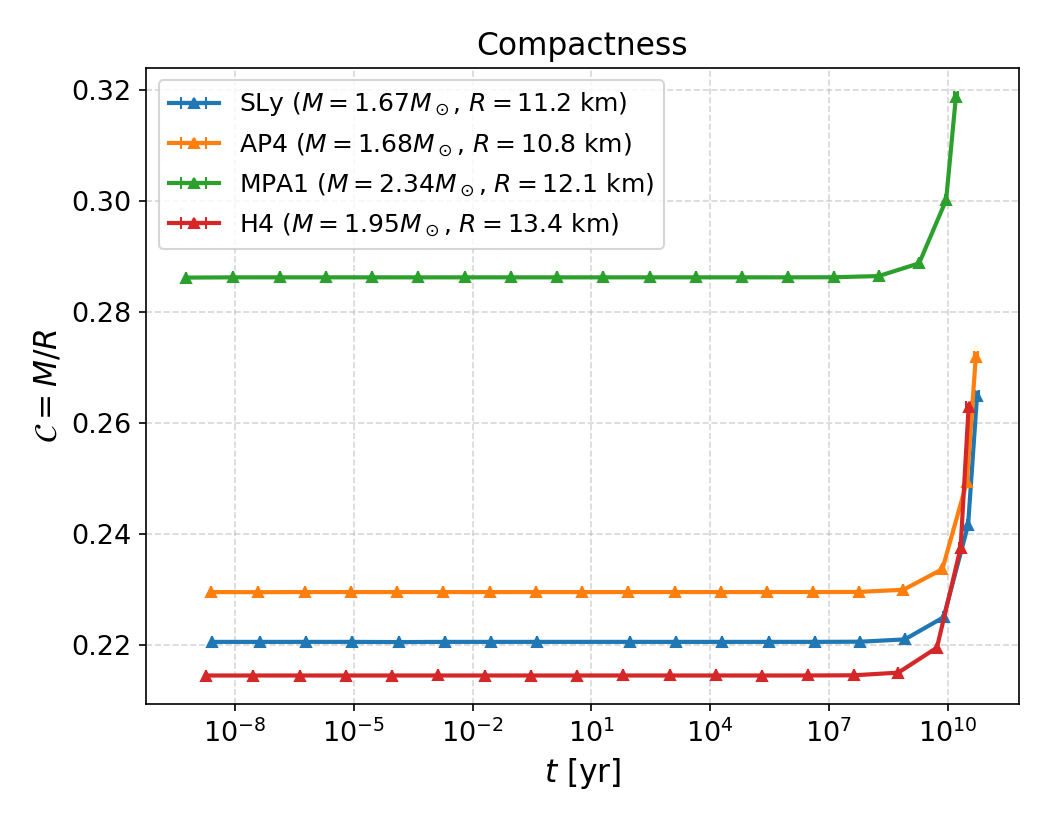}\\
    (a)~~~~~~~~~~~~~~~~~~~~~~~~~~~~~~~~~~~~~~~~~~~~~~~~(b)~~~~~~~~~~~~~~~~~~~~~~~~~~~~~~~~~~~~~~~~~~~~~~~~(c)\\
    \includegraphics[width=0.33\linewidth]{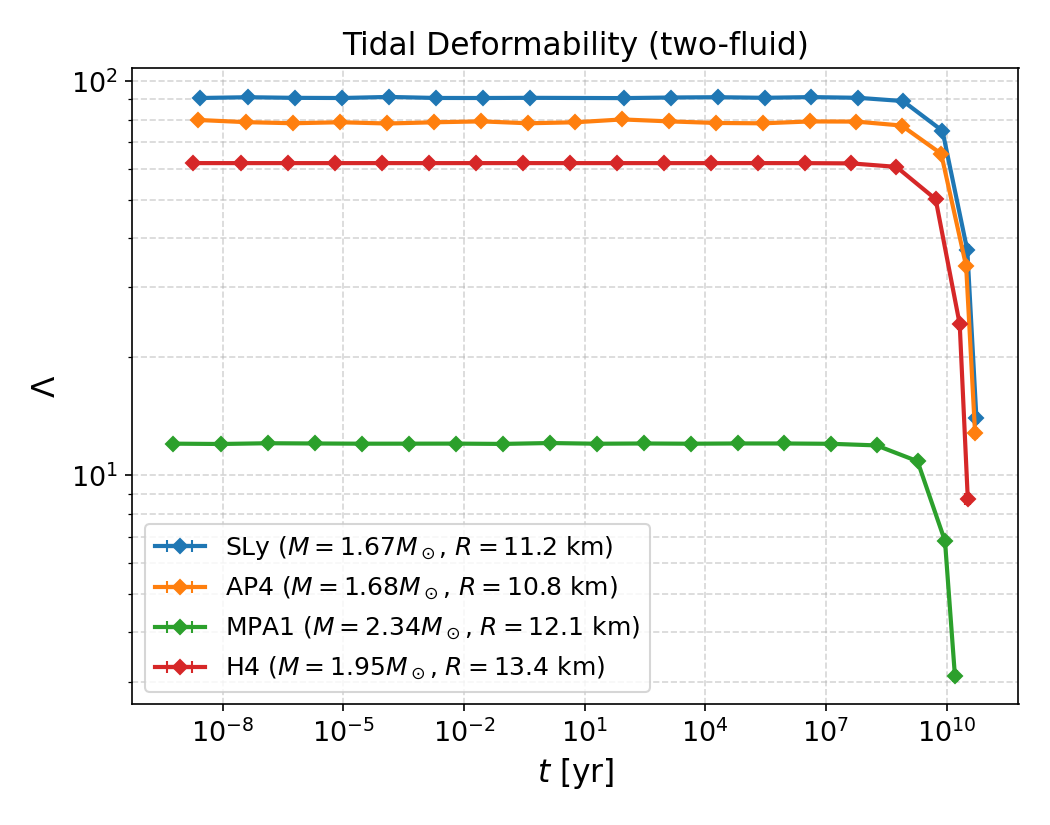}\includegraphics[width=0.33\linewidth]{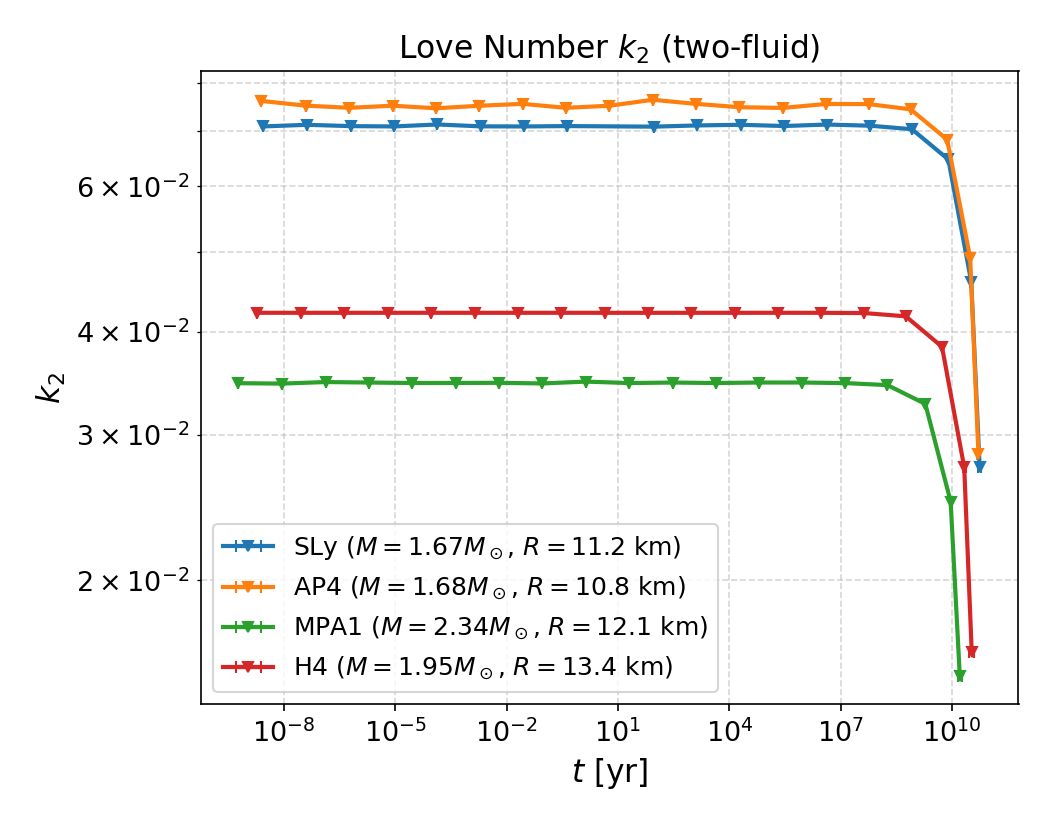}\includegraphics[width=0.33\linewidth]{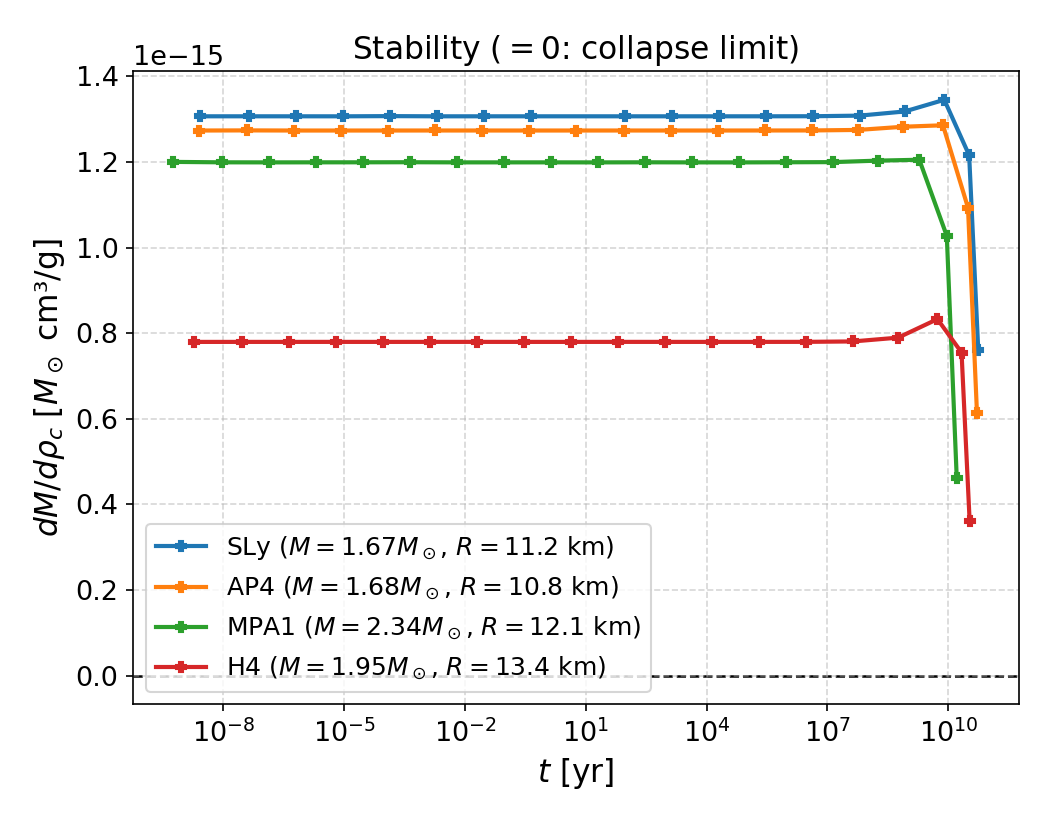}\\
    (d)~~~~~~~~~~~~~~~~~~~~~~~~~~~~~~~~~~~~~~~~~~~~~~~~(e)~~~~~~~~~~~~~~~~~~~~~~~~~~~~~~~~~~~~~~~~~~~~~~~~(f)\\
    \caption{
Comparison of the DM induced structural evolution for different NS EoS models (SLy, AP4, MPA1, and H4) under the maximized capture scenario with \(\rho_{\rm halo}=2.2\times10^9\,M_\odot/\mathrm{pc}^3\) and \(v_{\rm ns}=30\,\mathrm{km/s}\). The time evolution of (a) the accumulated DM particle number \(N_\chi\), (b) the DM distribution radius \(R_{\rm DM}\), (c) the stellar compactness \(C\), (d) the tidal deformability \(\Lambda\), (e) the Love number \(k_2\), and (f) the stability indicator \(dM_{\rm tot}/d\rho_c\). Although the quantitative responses differ among the EoS models, the overall evolution remains qualitatively similar.
}
    \label{figEOSdense}
\end{figure*}
\end{widetext}

From Figs.~\ref{figEOS} and \ref{figEOSdense}, we find that the effects of DM accumulation are qualitatively consistent across all EoS models. Softer EoS models exhibit slightly larger structural modifications owing to their higher compactness, whereas stiffer EoS models respond more weakly. Nevertheless, the overall differences among the EoS models remain modest, even in the maximized capture scenario.

\subsection{Dependence on the Initial Compactness}\label{subsecdependIC}

The DM capture rate depends sensitively on the spacetime geometry of the NS through the metric function \(A(r)\) in Eq.~\eqref{eqcap}. Consequently, NSs with different initial compactnesses can exhibit different capture efficiencies and therefore different long term DM accumulation histories.

To investigate this dependence, we compare equilibrium sequences constructed from NSs with different initial central baryon densities \(\rho_{c,0}\), which correspond to different initial compactnesses. Throughout this section, the EoS is fixed to the H4 model, and only the initial stellar compactness is varied. This setup isolates the influence of the background stellar geometry while minimizing uncertainties associated with the nuclear EoS.

Specifically, we consider the sequence of central densities
\begin{align}
    \rho_{c,0}
    =
    (1.05,\ 1.10,\ 1.15,\ 1.20,\ 1.30)
    \times10^{15}\,\mathrm{g\,cm^{-3}},
\end{align}
which generate NS configurations with progressively larger compactness.

For each configuration, we reconstruct the long term DM accumulation history and quantify how the capture efficiency and the resulting structural evolution depend on the initial compactness.

\begin{table}[htp!]
    \centering
    \scalebox{1.0}{
    \begin{tabular}{l|c|c|c|c|c}
    \hline\hline
    Star & $\rho_c/{\rm g\cdot cm^{-3}}$ & $M_{\rm ns}/M_{\odot}$ & $R_{\rm ns}/\mathrm{km}$ & $C$ & $M_{\rm tot}(t_H)/M_{\odot}$\\
    \hline\hline
    1 & $1.05\times10^{15}$ & $1.447$ & $12.091$ & $0.177$ & $1.458$\\
    \hline
    2 & $1.10\times10^{15}$ & $1.495$ & $11.953$ & $0.185$ & $1.505$\\
    \hline
    3 & $1.15\times10^{15}$ & $1.541$ & $11.822$ & $0.193$ & $1.550$\\
    \hline
    4 & $1.20\times10^{15}$ & $1.585$ & $11.697$ & $0.200$ & $1.602$\\
    \hline
    5 & $1.30\times10^{15}$ & $1.663$ & $11.459$ & $0.214$ & $1.679$\\
    \hline\hline
    \end{tabular}}
    \caption{
Representative H4 NS configurations with different initial central densities and corresponding compactnesses. Here \(M_{\rm ns}\) denotes the initial NS mass before DM accumulation, while \(M_{\rm tot}(t_H)\) denotes the total stellar mass after a Hubble time of DM accumulation.
}
    \label{tabcompareIC}
\end{table}

Table~\ref{tabcompareIC} gives the baseline NS configurations considered in this section. Increasing the initial central density produces progressively more compact stars with smaller radii and larger gravitational masses, allowing us to investigate how the long term effects of DM accumulation depend on the initial compactness.

\begin{figure}[htp!]
    \centering
    \includegraphics[width=0.8\linewidth]{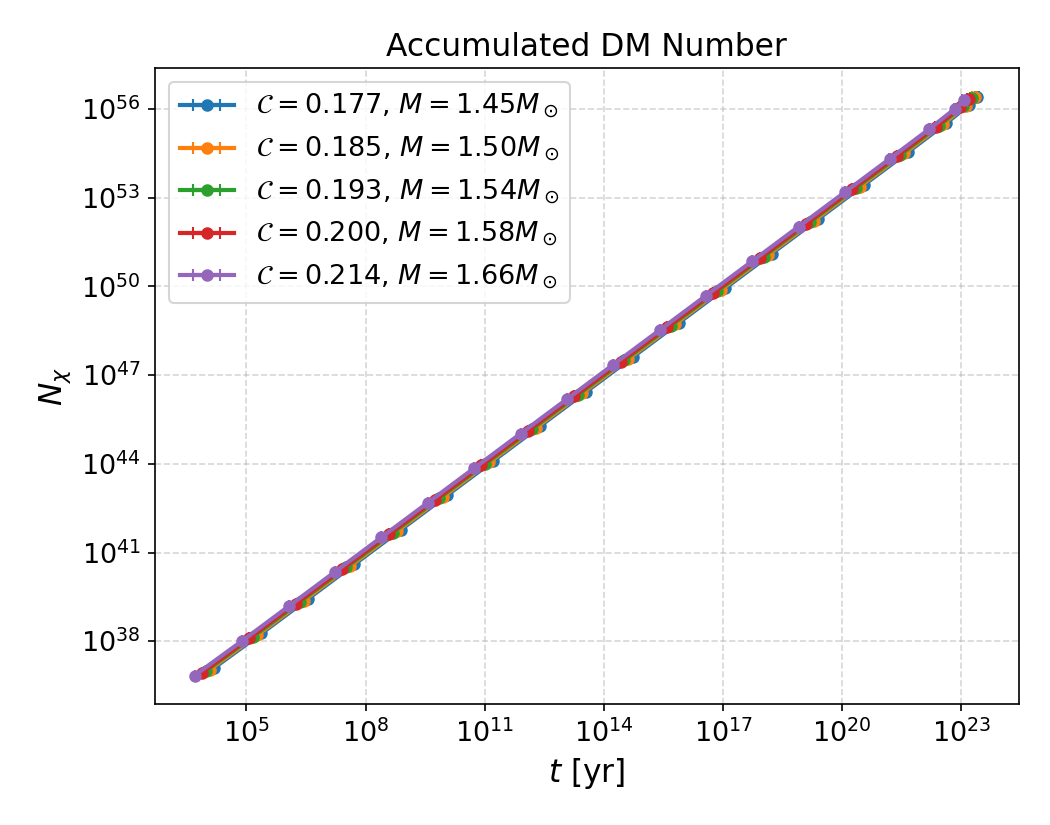}\\
    ~~~~(a)\\
    \includegraphics[width=0.8\linewidth]{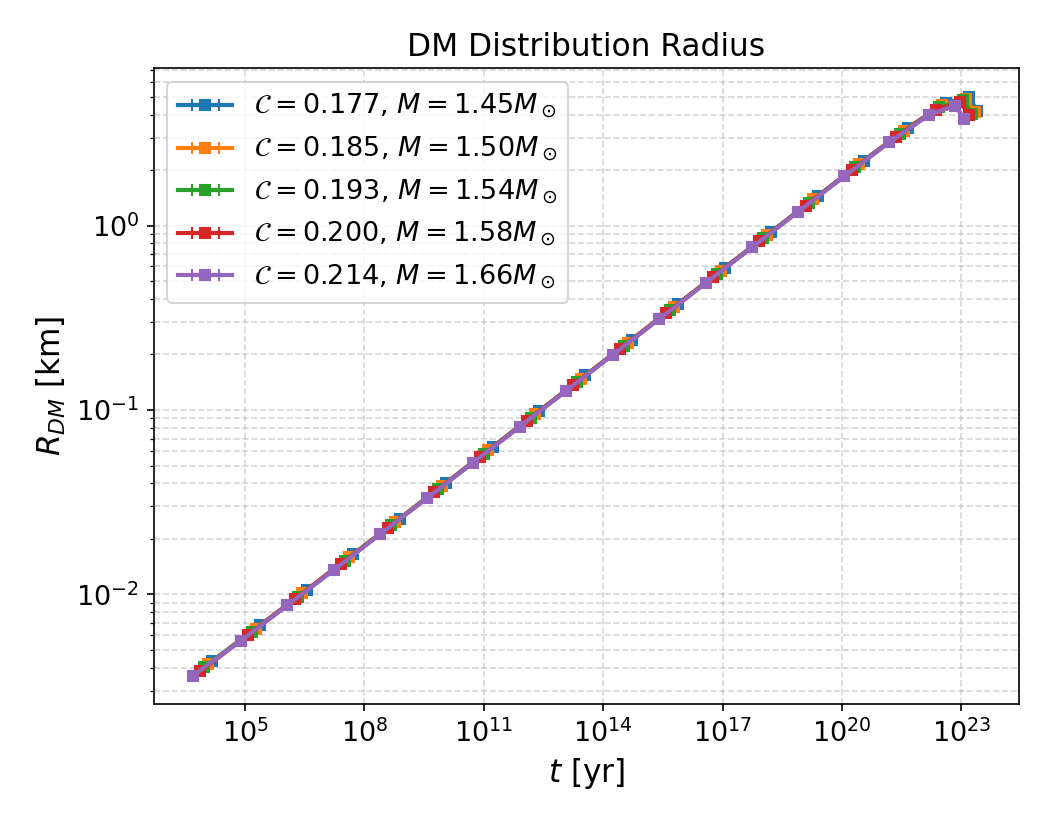}\\
    ~~~~(b)\\
    \caption{
Evolution of NS properties for configurations with different initial compactness under the canonical Galactic background DM density. (a) shows the time evolution of the accumulated DM particle number \(N_\chi\), (b) shows the evolution of the DM distribution radius \(R_{\rm DM}\).
}
    \label{figcompactnesscomparison1}
\end{figure}

\begin{figure}[htp!]
    \centering
    \includegraphics[width=0.8\linewidth]{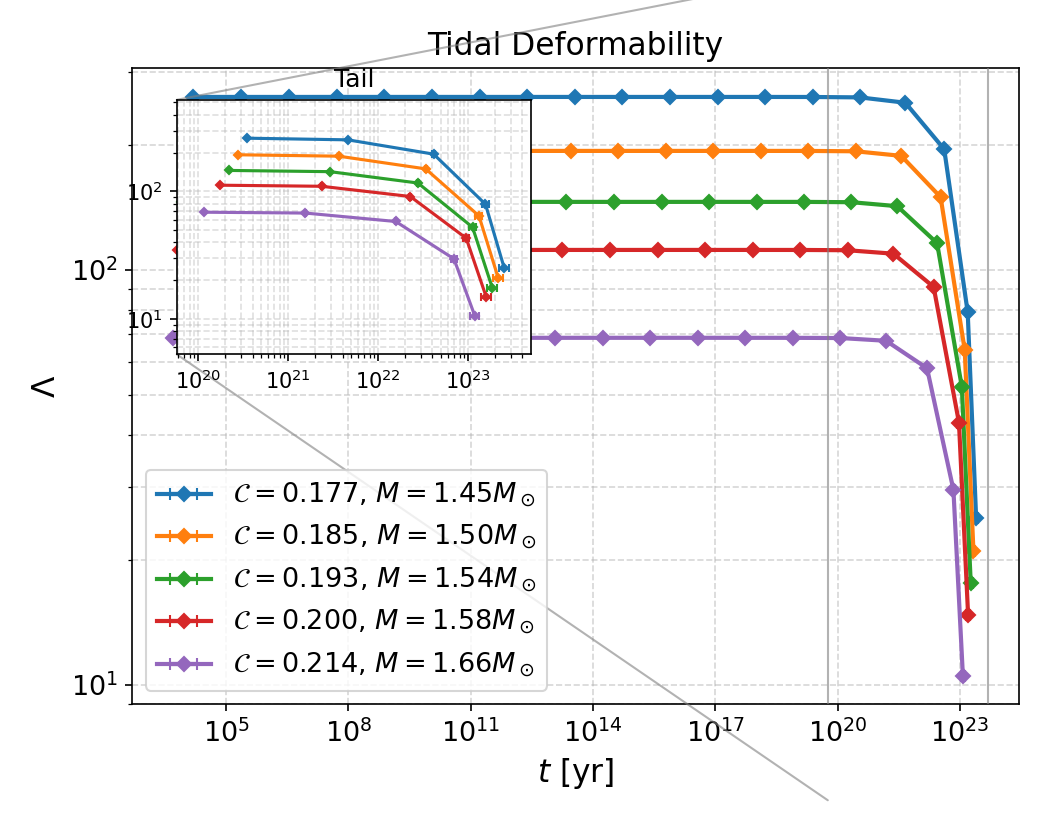}\\
    ~~~~(a)\\
    \includegraphics[width=0.8\linewidth]{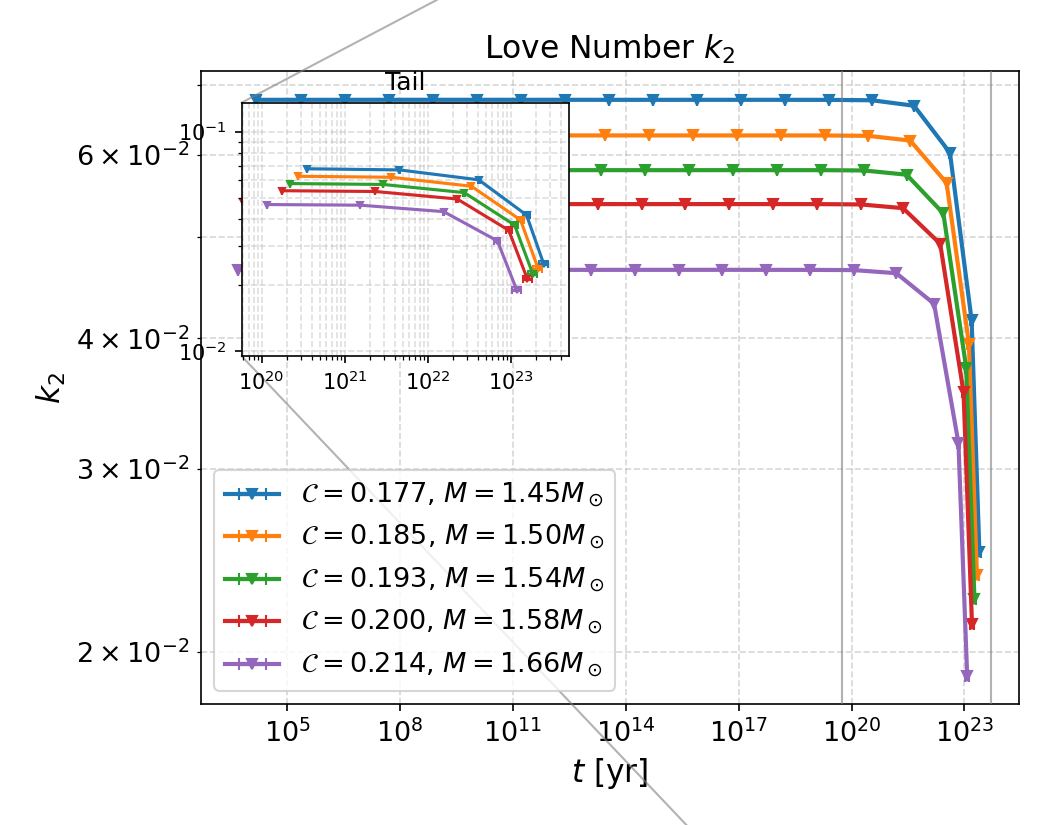}\\
    ~~~~(b)\\
    \caption{
Evolution of NS properties for configurations with different initial compactness under the canonical Galactic background DM density. (a) shows the evolution of the tidal deformability \(\Lambda\), (b) shows the evolution of the Love number \(k_2\). More compact NSs exhibit smaller initial tidal deformabilities and stronger suppression of the tidal response under DM accumulation.
}
    \label{figcompactnesscomparison2}
\end{figure}

\begin{figure}[htp!]
    \centering
    \includegraphics[width=0.8\linewidth]{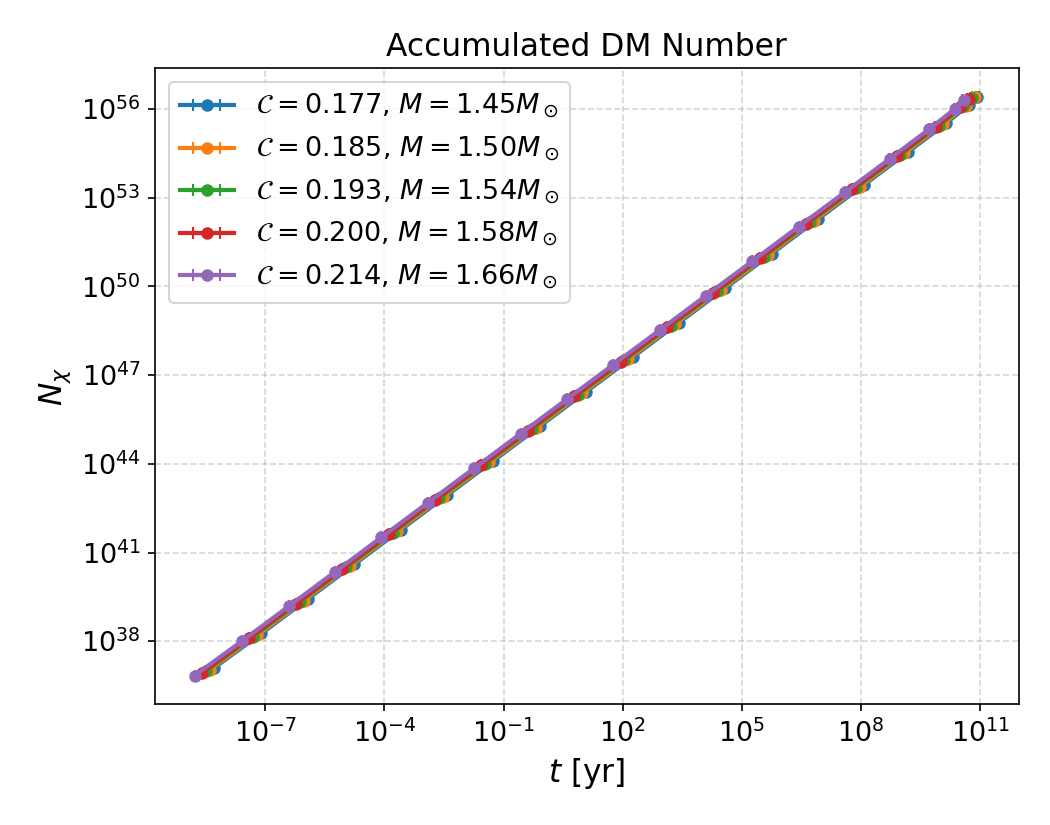}\\
    ~~~~(a)\\
    \includegraphics[width=0.8\linewidth]{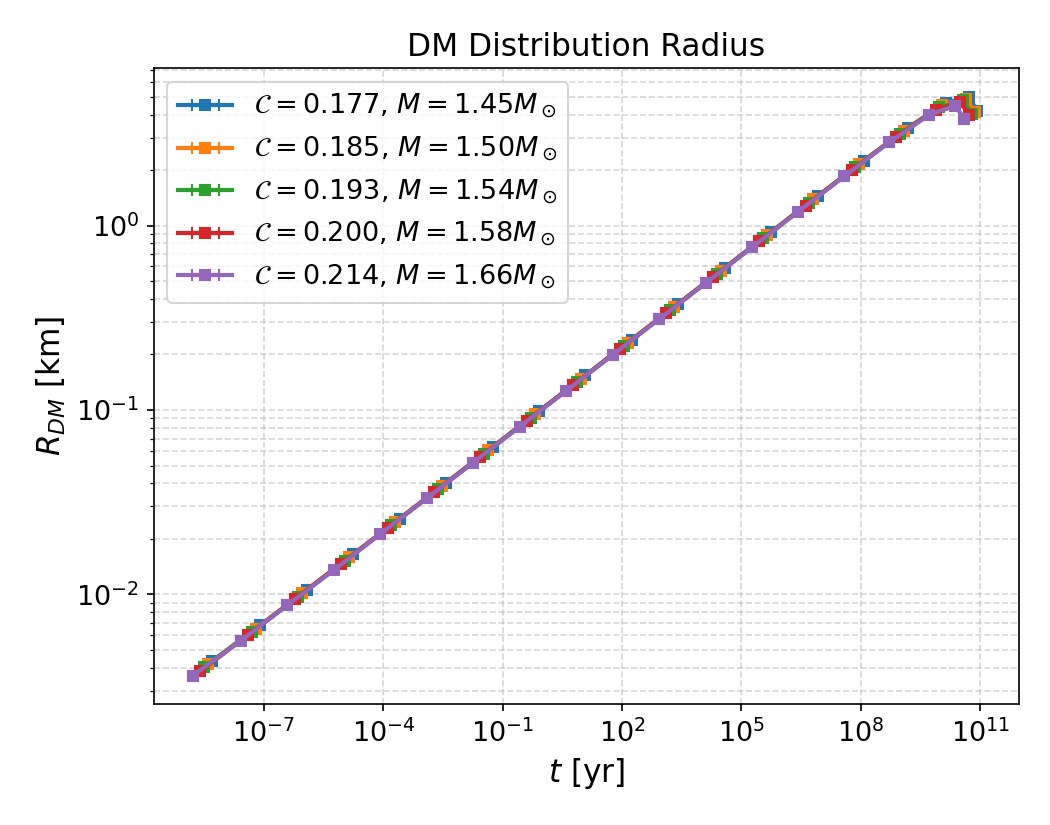}\\
    ~~~~(b)\\
    \caption{
Evolution of NS properties for configurations with different initial compactness under the maximized capture scenario with dense DM spikes. (a) shows the time evolution of the accumulated DM particle number \(N_\chi\), (b) shows the evolution of the DM distribution radius \(R_{\rm DM}\).
}
    \label{figcompactnesscomparisondense1}
\end{figure}

\begin{figure}[htp!]
    \centering
    \includegraphics[width=0.8\linewidth]{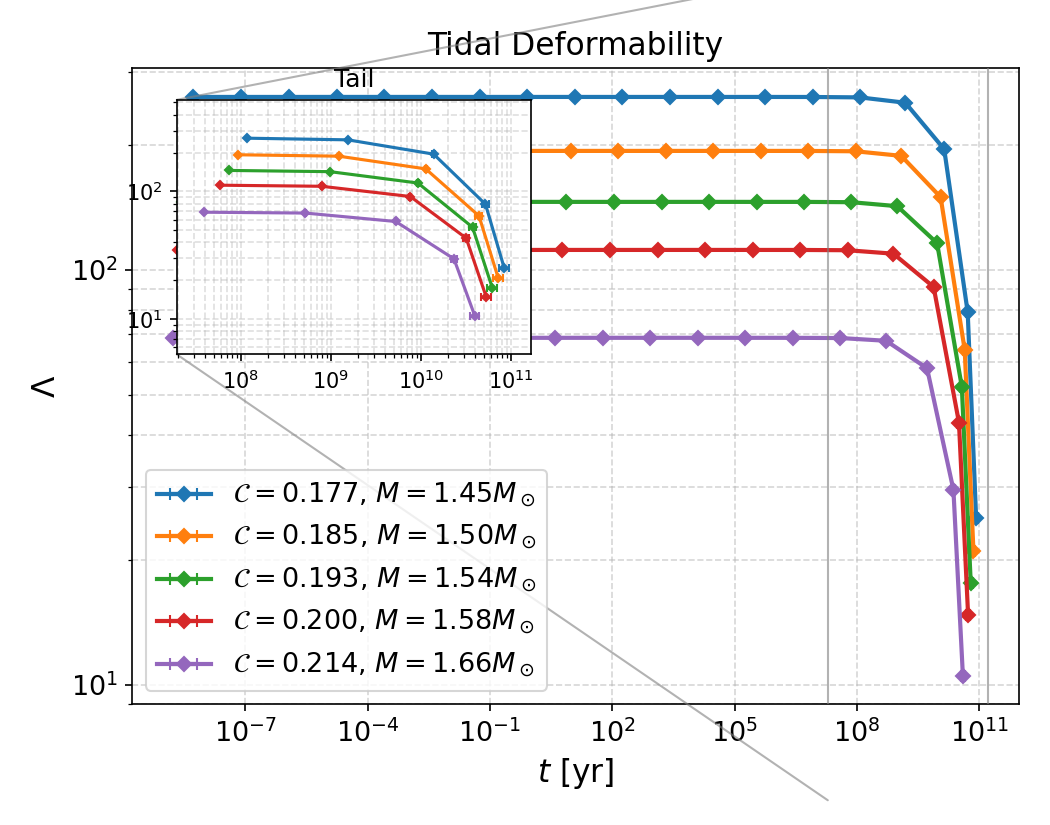}\\
    ~~~~(a)\\
    \includegraphics[width=0.8\linewidth]{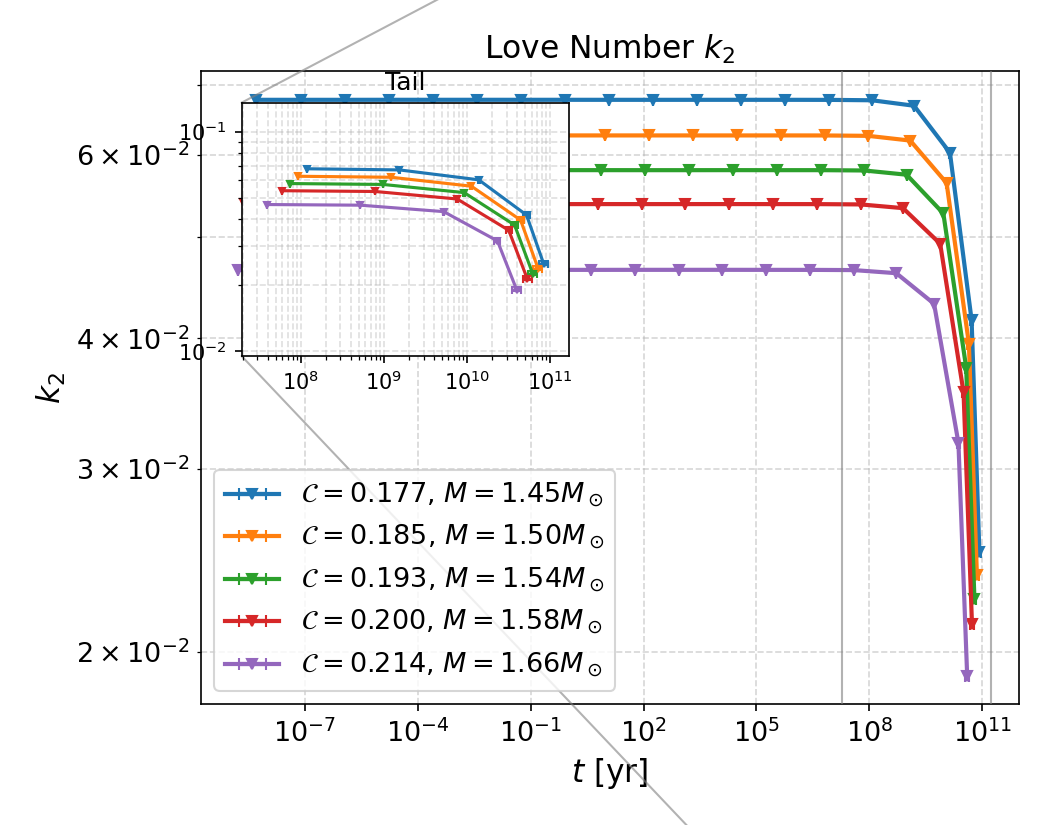}\\
    ~~~~(b)\\
    \caption{
Evolution of NS properties for configurations with different initial compactness under the maximized capture scenario with dense DM spikes. (a) shows the evolution of the tidal deformability \(\Lambda\), (b) shows the evolution of the Love number \(k_2\). More compact NSs exhibit smaller initial tidal deformabilities and stronger suppression of the tidal response under DM accumulation.
}
    \label{figcompactnesscomparisondense2}
\end{figure}

The evolution of the Love number and tidal deformability shown in Figs.~\ref{figcompactnesscomparison2} and \ref{figcompactnesscomparisondense2} illustrates the dependence of the NS tidal response on the initial compactness. More compact NSs have smaller initial values of \(k_2\) and \(\Lambda\) and capture DM more efficiently owing to their deeper gravitational potentials and stronger relativistic gravitational focusing.

Since the tidal deformability scales as
\begin{align}
    \Lambda \propto k_2 C_{\rm NS}^{-5},
\end{align}
its strong dependence on the stellar compactness amplifies the effect of structural variations. Nevertheless, the overall modifications remain small, indicating that DM accumulation primarily enhances pre-existing differences among NS configurations rather than introducing qualitatively new tidal behavior.

For Galactic background DM densities, the tidal response evolves only negligibly over a Hubble time, with both the Love number and tidal deformability remaining nearly unchanged. Appreciable deviations appear only on much longer timescales,
\begin{align}
    t\sim10^{20}\,\mathrm{yr},
\end{align}
well beyond realistic astrophysical evolution.

In contrast, under the maximized capture scenario, the tidal response exhibits noticeable evolution within a Hubble time, with
\begin{align}
    |\Delta k_2|
    \sim
    10^{-2},
    \qquad
    |\Delta\Lambda|
    \sim
    10^{1},
\end{align}
respectively.

As discussed in Sec.~\ref{subseck2}, appreciable modifications of the tidal response occur only in deliberately extreme DM environments and remain difficult to detect with current gravitational wave observations.

Our results also indicate that more compact NSs can accumulate DM slightly more efficiently under identical environmental conditions due to their deeper gravitational potentials and stronger relativistic gravitational focusing. This effect may have implications for binary systems. If both NSs are embedded in the same ambient DM environment, the more compact component is expected to capture DM at a slightly faster rate. For a fixed EoS, the more massive NS is typically more compact. Therefore, for a mass ratio
\begin{align}
    q=\frac{M_1}{M_2}<1,
\end{align}
with \(M_1<M_2\), the heavier component may gradually accumulate a larger DM fraction over astrophysical timescales.

Even in the maximized capture scenario, the resulting evolution of the mass ratio remains modest over a Hubble time. These results therefore suggest that capture driven DM accumulation is unlikely to generate significant mass asymmetries in binary NS systems, although the mechanism may provide a useful phenomenological framework for future studies of DM effects in compact binaries.

\subsection{Dependence on the Dark Matter Particle Mass}

The DM particle mass plays a central role in determining the efficiency of the capture process and therefore strongly affects the long term accumulation history inside the NS. In particular, the capture rate depends explicitly on \(m_\chi\) through both the incident DM number density and the scattering kinematics appearing in Eq.~\eqref{eqcap}. Since lighter DM particles are generally captured more efficiently, varying \(m_\chi\) can significantly change the accumulated DM fraction and the resulting modification of the NS structure.

\begin{widetext}
    \begin{figure*}[htp!]
    \centering
    \includegraphics[width=1.0\linewidth]{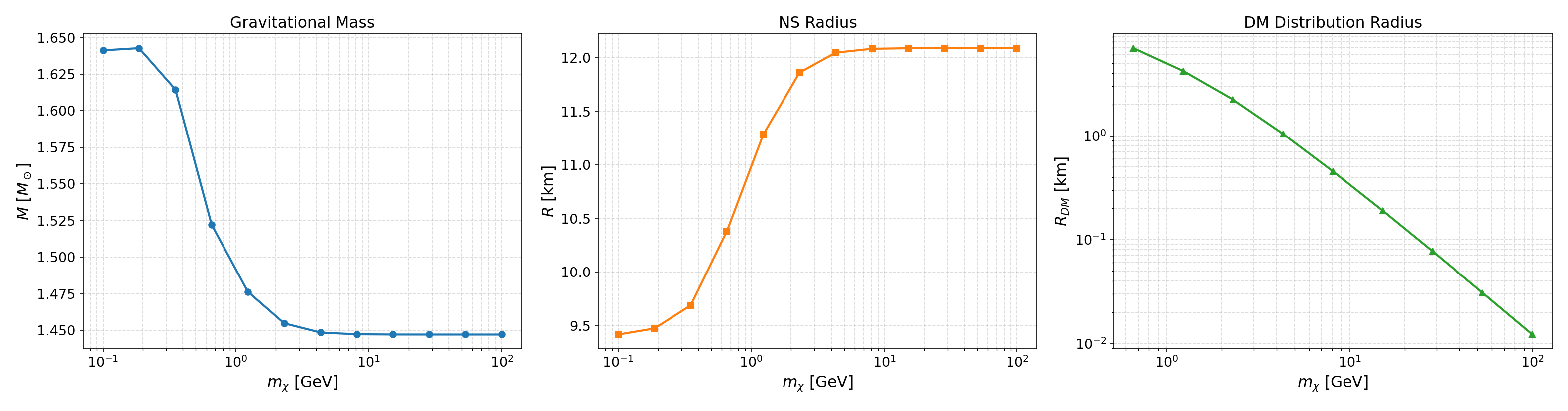}\\
    (a)~~~~~~~~~~~~~~~~~~~~~~~~~~~~~~~~~~~~~~~~~~~~~~~~(b)~~~~~~~~~~~~~~~~~~~~~~~~~~~~~~~~~~~~~~~~~~~~~~~~(c)\\
    \includegraphics[width=1.0\linewidth]{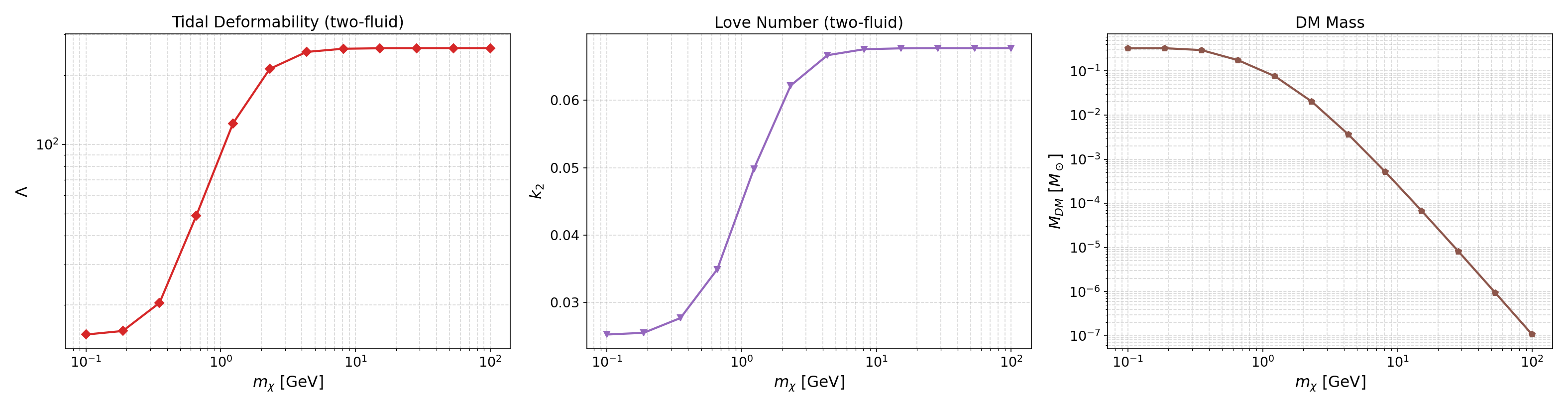}\\
    (d)~~~~~~~~~~~~~~~~~~~~~~~~~~~~~~~~~~~~~~~~~~~~~~~~(e)~~~~~~~~~~~~~~~~~~~~~~~~~~~~~~~~~~~~~~~~~~~~~~~~(f)\\
    \caption{
Dependence of NS and DM properties on the DM particle mass \(m_{\chi}\) over the range \(0.1\!-\!100\,\mathrm{GeV}\). (a) the total stellar mass \(M_{\rm tot}\), (b) the NS radius \(R_{\rm ns}\), (c) the DM distribution radius \(R_{\rm DM}\), (d) the tidal deformability \(\Lambda\), (e) the Love number \(k_2\), and (f) the accumulated DM mass \(M_{\chi}\). Lighter DM particles produce larger accumulated DM fractions and correspondingly stronger structural modifications.
}
    \label{figmchi_scan}
\end{figure*}
\end{widetext}

To investigate this dependence, we vary the DM particle mass over the range
\[
0.1\!-\!100\,\mathrm{GeV},
\]
and compute the corresponding NS and DM properties. The results are shown in Fig.~\ref{figmchi_scan}.

As shown in Fig.~\ref{figmchi_scan}, lighter DM particles produce larger accumulated DM masses owing to the approximate scaling
\begin{align}
    C_{\rm cap}\propto m_{\chi}^{-1},
\end{align}
which increases the capture efficiency for a fixed ambient DM density.

As a result, lower mass DM produces slightly larger modifications of the NS structure and tidal response. Nevertheless, these effects remain modest throughout the mass range considered in this work, consistent with our previous conclusions.

\section{Conclusion}\label{secconclusion}

In this work, we developed a self-consistent framework to investigate the long term accumulation of asymmetric fermionic DM inside NSs and its impact on NS structure and gravitational wave observables. By combining a piecewise polytropic NS EoS with the coupled two fluid TOV equations and a time dependent DM capture formalism, we reconstructed the quasi-static evolution of both the accumulated DM component and the stellar structure over cosmological timescales.

Within this framework, we investigated how the accumulation history and the resulting structural evolution depend on the ambient DM environment, the initial NS compactness, the nuclear EoS, and the DM particle mass. We further quantified the evolution of key observable quantities, including the stellar compactness, Love number, and tidal deformability, and examined the stability of the system against gravitational collapse.

Our results show that the amount of DM accumulated inside NSs remains strongly limited, even under deliberately optimized capture conditions involving light asymmetric fermionic DM embedded in dense DM spikes together with low NS velocities. Consequently, the accumulated DM component always remains gravitationally subdominant, and the resulting structural modifications remain modest over a Hubble time.

DM accumulation produces a gradual increase in the stellar compactness together with a suppression of the Love number and tidal deformability. Because the tidal deformability scales approximately as \(\Lambda \propto k_2 C^{-5}\), relatively small changes in the compactness can generate amplified variations in \(\Lambda\). However, noticeable evolution of the tidal response occurs only in deliberately extreme Galactic center environments and over cosmological timescales comparable to a Hubble time. For realistic Galactic or cluster DM densities, the resulting deviations remain far below the sensitivity of current or near future gravitational wave and electromagnetic observations.

We also find that NSs with larger initial compactness generally capture DM more efficiently due to their deeper gravitational potentials and stronger relativistic focusing. In binary systems, this differential capture induces only a weak mass ratio drift because the accumulated DM fraction remains small throughout the evolution.

Overall, our analysis indicates that asymmetric DM is unlikely to generate large observable modifications of NS structure through capture driven accumulation alone. Nevertheless, the framework developed here provides a self-consistent method for quantifying these effects and establishing conservative upper bounds on DM induced structural modifications in compact stars.

Possible extensions of this work include self-interacting or bosonic DM, rotation, magnetic fields, finite temperature effects, and dynamically evolving astrophysical environments. It would also be interesting to investigate whether alternative dark sector interactions or more efficient capture mechanisms could produce observable signatures in future gravitational wave observations of compact binary systems.

\vspace{2cm}

\bibliographystyle{apsrev}
\bibliography{references}
\end{document}